\documentstyle[prd,preprint,aps]{revtex}

\begin{document}
\draft

\title{Dynamics of Gravitational Waves in 3D: Formulations,
Methods, and Tests}

\author{Peter Anninos${}^{(1)}$, Joan Mass\'o${}^{(1,2)}$, 
Edward Seidel${}^{(1,3)}$, Wai-Mo Suen${}^{(1,4,5)}$, and Malcolm 
Tobias${}^{(4)}$}

\address{${}^{(1)}$ National Center for Supercomputing Applications,
605 E. Springfield Ave., Champaign, Illinois 61820}

\address{${}^{(2)}$ Department of Physics,
Universitat de les Illes Balears, Palma de Mallorca E-07071, Spain}

\address{${}^{(3)}$ Department of Physics,
University of Illinois at Urbana-Champaign}

\address{${}^{(4)}$McDonnell Center for the Space Sciences,
Department of Physics,
Washington University, St. Louis, Missouri 63130}

\address{${}^{(5)}$Physics Department, Chinese University of Hong
Kong, Shatin, Hong Kong.}

\date{\today}
\maketitle
\begin{abstract}

The dynamics of gravitational waves is investigated in
full 3+1 dimensional numerical relativity, emphasizing 
the difficulties that one might encounter
in numerical evolutions, particularly those 
arising from non-linearities and gauge degrees of freedom.
Using gravitational waves with amplitudes low enough that 
one has a good understanding of the physics involved, but large enough 
to enable non-linear effects to emerge, we study the coupling between 
numerical errors, coordinate effects, and the nonlinearities of the 
theory.  We discuss the various strategies used in identifying specific
features of the evolution.  We show the importance of the flexibility 
of being able to use different numerical schemes, different slicing
conditions, different formulations of the Einstein equations 
(standard ADM vs. first order hyperbolic), and different sets of 
equations (linearized vs. full Einstein equations).  A
non-linear scalar field equation is presented
which captures some properties of the full 
Einstein equations, and has been useful in our understanding of the 
coupling between finite differencing errors and non-linearites.  We 
present a set of monitoring devices which have been crucial in our 
studying of the waves, including Riemann invariants, pseudo-energy 
momentum tensor, hamiltonian constraint violation, and fourier 
spectrum analysis.

\end{abstract}


\narrowtext

\section{Introduction}
\label{introduction}

This paper is the first in a series of papers in which we numerically
study gravitational waves in 3+1 dimensions.  The systems studied
range from weak gravitational waves with various symmetries, to fully
general and highly nonlinear waves.  
We study the dynamical evolutions of the waves
and the interactions between waves.  That is, we investigate the
dynamics of spacetime in its pure (vacuum) form, a subject that is
important for theoretical, observational, and technical reasons.  This
area of research is for the most part uncharted territory due to its
mathematical complexity and the need for large scale computational
resources that have not been available previously.  The general
behavior of three-dimensional (3D) strong gravitational waves,
including for example gravitational geons and the formation of
singularities, is unknown.  Previous analytic and numerical work on
pure gravitational wave spacetimes, done in one or two spatial
dimensions, has led to many interesting results, such as the formation
of singularities from colliding plane
waves~\cite{Khan71,Tipler80,Matzner84,Yurtsever88a,Yurtsever88b} or
the formation of black holes by imploding axisymmetric gravitational
waves~\cite{Abrahams92b} and the existence of critical behavior in
such systems~\cite{Abrahams93a}.  These discoveries raise interesting
questions about waves in more general 3D spacetimes.  Yurtsever has
proposed conjectures concerning the criteria for the formation of
singularities from wave packets with finite extension in all 3
directions~\cite{Yurtsever88a,Yurtsever88b}.  These conjectures,
together with the global structure and local behavior of the
singularities so formed, if indeed they can be formed, remains to be
investigated.  These questions call for a full 3D study.

Gravitational waves are also about to open up a fundamentally new area
of astronomical observation: gravitational wave astronomy.  A new
generation of interferometric~\cite{LIGO} and bar
detectors~\cite{Johnson93} should detect waves for the first time near
the turn of the century.  Even though any observed waves are expected
to have weakened by the time they reach Earth, they are likely to have
been generated in regions with strong, highly dynamical and nonlinear
gravitational fields.  It is therefore essential to be able to study
waves accurately in both the strong and weak field regimes, as well as
the long term secular behavior in the transitory intermediate regimes.
The study of pure wave spacetimes will aid us in developing numerical 
codes to study all three regimes with confidence in the numerical results.

These pure wave studies compliment our program to compute the
evolution and the radiation from the coalescence of two black holes in
decaying orbits.  Because black hole and gravitational wave
systems each present their own set of technical difficulties, we first
study black holes and waves separately, and then combine them after
the problematics of each system are identified and understood.  In a
separate paper~\cite{Anninos94c} we have presented results for a pure
single black hole spacetime (i.e. Schwarzschild) evolved in
three-dimensional Cartesian coordinates, with essentially the same
basic code as used here.  In future papers
we will present results from evolutions of distorted black holes,
including both gravitational waves and black holes.

In this first paper we focus on examining the difficulties one 
encounters in evolving relatively low amplitude 3D gravitational waves 
in Cartesian coordinates, and on the strategies we developed to solve 
those problems.  We begin with low amplitude waves, as one has better 
physical understanding of what should be happening in such cases.  
When the amplitude is very low, the evolution is linear and nothing 
interesting happens.  What is more interesting is waves in the 
``near-linear'' regime, the meaning of which will become clearer 
throughout the paper.  Basically, we mean waves which show some 
nonlinear transient or secular effects that can be observed in our 
numerical study within the limit of the accuracy and time scale of the 
evolution that we can currently achieve.  These effects could be due 
to (i) numerical errors (finite differencing errors) coupled with 
nonlinearity, (ii) coordinate effects due to nonlinearity, or (iii) 
nonlinear physics.  [Of course there is also a (iv) that we have 
invested a lot of effort in making sure of its absence (an effort which 
is not discussed here), namely, coding errors].  In this first paper, 
our aim is to study (i) and (ii) instead of (iii).  We find that there 
are indeed cases for which (i) and (ii) give rise to interesting 
features in the evolution, but have negligible nonlinear physical 
effects.

It is nontrivial to distinguish whether a feature is due to (i), (ii), 
or (iii).  To make the distinction between these effects, we have 
implemented many monitors of the evolution, e.g.  Hamiltonian 
constraint, pseudo-energy-momentum tensor, curvature components and 
curvature invariants.  We have the options of using different gauge 
and slicing conditions, different boundary conditions, different 
finite differencing schemes, with different orders of finite 
differencing.  In addition to the codes that evolve the full 3-D 
nonlinear Einstein equations, we have developed other evolution 
codes for comparison, e.g., codes that evolve the linearized Einstein
equations, and  
codes that evolve a scalar field equation that captures important 
features of the full Einstein equations.  Most noteworthy is that we 
have developed two completely independent codes that are based on two 
very different analytic formulations of the full Einstein evolution
equations.   
All simulations presented in this paper were run with both codes, and 
the results were compared in detail.  It is important to point out 
that the two codes will not produce identical results.  One code is 
based on a particular gauge choice where that gauge condition is
assumed in the evolution equations, whereas the other code has the
evolution equations in their completely general from.  When a
gauge is chosen it can only be kept to numerical error.

The first of these fully nonlinear codes, which we call the ``G''
(for General) code, is based on the standard 3+1 ADM~\cite{Arnowitt62}
approach to numerical relativity.  It has been written in a fully
general way, without specializing the equations to any lapse or shift
condition, and without any restrictions on symmetry or initial data.
The second code, which we call the ``H'' (for Harmonic) code, is based
on the first order, flux-conservative, hyperbolic formulation of the
Einstein equations developed by Bona and Mass\'o~\cite{Bona89,Bona92}.
Different finite differencing and evolution schemes have been
incorporated into both codes, as well as linearized versions of
both formulations.  All these different codes and options were
essential in enabling us to sort out the effects (i)-(iv) mentioned
above.

We discuss three types of testbeds in this paper.  The first test we
consider is a single plane symmetric wave packet, propagating in some
arbitrary direction.  This problem allows us to compare the dispersive
and dissipative properties of the codes for waves propagating in
different directions in the 3D Cartesian grid, and the resolution
needed for a given desired accuracy.

The second type of test we consider is the collision of weak plane
wave packets.  The focus here is on an effect caused by a coupling
between finite differencing errors and the nonlinearity of the
evolution equations.  It manifests itself as a drifting of the metric
function in a region where the wave packet has crossed.  We discuss in
detail how the cause of this drift can be identified.  We develop a
scalar field equation which captures important features of the
nonlinear evolution of the Einstein equations.  Testbeds done with
this equation have been crucial in this analysis.  We propose that this
scalar equation be used as a standard testbed for the numerical study of
gravitational waves.

The third type of testbed is an imploding-exploding combination
of quadrupole wave 
packets~\cite{Eppley79,Teukolsky82}.  Besides analyzing the accuracy
of the numerical evolution, the focus here is on the coupling between
the motion of the coordinates and the
nonlinearity of the Einstein equations.  With geodesic slicing, this
coupling manifests itself as a ``dipping'' of some metric functions at
the center of the symmetry, at a time long after the implode-explode
process. We study at what amplitude this phenomenon
becomes observable.  We report on the analysis carried out in
confirming that this behavior is due solely to coordinate motion
instead of truly nonlinear physics.

For all three types of testbeds, we have studied the evolution of
initial data sets which satisfy the constraint equations to linear
order, and for the third testbed, data that completely satisfies the
constraints, obtained through the York's formalism \cite{York79}.  We
have checked that the two kinds of initial data basically lead to the
same kinds of evolution for the low amplitude waves studied in this paper,
hence we do not discuss the two cases separately unless otherwise
mentioned.  Throughout the paper we restrict ourselves to time
symmetric initial data for simplicity when solving for the initial
value problem, which is not our major concern in this paper.

In this paper we use the convention of~\cite{Misner73}, in
which $c=1$ and, as we are studying vacuum spacetimes, $G$ does not enter.  The
system has no intrinsic length scales except those set by the waves,
e.g. wavelength.

This paper is separated into the following sections:
Section~\ref{formalisms} reviews the two different codes we have
developed, which are based on the two different analytic formulations
of the Einstein equations.  We also discuss the numerical methods used
in these two codes.  The different tests and comparisons of our codes
are presented in Sections~\ref{III}-\ref{VI}.  Section~\ref{III} is on
planewave packets. Section~\ref{IV} is on colliding packets.
Section~\ref{V} discusses a nonlinear scalar field equation that is
useful in analyzing the nonlinearity of the Einstein evolution
equations.  Section~\ref{VI} is on imploding-exploding quadrupole
waves. Section~\ref{VII} is a brief discussion and conclusion.

\section{Basic Formalisms and Numerical Methods}
\label{formalisms}

\subsection{The Fully Nonlinear 3D Codes}
\label{nonlinear}

We have developed two independent 3D codes to solve the fully
nonlinear set of Einstein equations.  As all tests presented in this
paper are performed with both codes, this approach allows us to study
systematically the effect of not only different numerical methods, of
which we have tested several, but also different mathematical
formulations of the equations.

\subsubsection{The ``G'' code}
\label{thegcode}

The first code we present is the ``G'' code, where G stands for general.  This
code uses the standard 3+1 ADM formulation of the Einstein equations.
It is general in the sense that it can be used with arbitrary slicing
and spatial coordinate conditions.
The general spacetime metric is of the form
\begin{equation}
ds^2 = (-\alpha^2 + \beta_i \beta^i) dt^2 + 2\beta_i dx^i dt
        +g_{ij} dx^i dx^j~,
\label{4metric}
\end{equation}
where $\alpha$ and $\beta^i$ are the lapse function and shift vector
respectively.  Although the vacuum ADM equations are given in many
papers, we again show them here so that one may compare them with a
second formulation discussed below:
\begin{eqnarray}
\partial_t g_{ij} &=& - 2 \alpha K_{ij}+\nabla_i \beta_j+\nabla_j
\beta_i
\label{dtgij} \\
\partial_t K_{ij} &=& -\nabla_i \nabla_j \alpha + \alpha \left(
R_{ij}+K\ K_{ij} -2 K_{im} K^m_j \right) \nonumber \\ &\ & + \beta^m
\nabla_m K_{ij}+K_{im} \nabla_j \beta^m+K_{mj} \nabla_i \beta^m~.
\label{dtkij}
\end{eqnarray}
Here $\nabla_i$ is the spatial covariant derivative, $R_{ij}$ is the
spatial Ricci tensor and $K$ is the trace of the extrinsic
curvature. While the code admits arbitrary kinematic conditions for
$\alpha$ and $\beta^i$, in this paper we report only on results
obtained with either geodesic slicing ($\alpha=1$), maximal slicing or
harmonic slicing for the lapse function, and zero shift vector.  
The maximal slicing lapse
\begin{equation}
\nabla^m \nabla_m \alpha = \alpha R~,
\end{equation}
is derived by taking the trace of (\ref{dtkij}) and setting
$K=\partial_t K = 0$.  
The harmonic slicing condition for the lapse is derived imposing the
harmonic condition on the time coordinate, leading to the evolution
equation 
\begin{equation}
\partial_t \alpha = -\alpha^2 K
\label{harmonic}
\end{equation}
where the initial value for the lapse is completely arbitrary.

It is also appropriate to introduce the Hamiltonian constraint:
\begin{equation}
h = R + K^2 - K_{ij} K^{ij} = 0~.
\label{ham}
\end{equation}
Although the evolution equations theoretically preserve the
Hamiltonian constraint in time, this is not generally so in
numerically constructed spacetimes. Discretization effects accumulate
over time, which can lead to violations of the Hamiltonian constraint.
The quantity $h$ defined in Eq.~(\ref{ham}) therefore offers a means
of monitoring errors introduced in the numerical evolution.

Equations (\ref{dtgij}) and (\ref{dtkij}) are expanded in a 3D
Cartesian coordinate system and coded in FORTRAN using MACSYMA scripts
written originally by David Hobill.  More details of this code are
provided in Ref.~\cite{Anninos94c}, where it was applied to black hole
spacetimes.

An important point to stress is that the equations have not been
specialized in any way.  All gauge degrees of freedom are left
general, so that any shift and lapse conditions may easily be imposed.
On the other hand, this implies that if a particular gauge choice is
used for the initial data (i.e., a diagonal form of the metric or a
traceless extrinsic curvature), the equations themselves are not
specialized to that gauge, and this allows for the possibility that
the gauge condition may not be strictly satisfied after some evolution
due to numerical errors.  We view it as an important strength of this
code, as it opens up the possibility of
investigating the stability of various gauge choices.

This code is sufficiently flexible that it allows different evolution
schemes to be implemented easily, and we have developed the following
two numerical schemes that are second order accurate in space and
time: a staggered leapfrog with half time step extrapolation, and a
``MacCormack-like'' predictor-corrector method.  An essential
difference between them is that in the MacCormack scheme, all
quantities are centered on the same time slices at all times, and
therefore no extrapolations or averages are needed to get quantities
that are properly centered.  The leapfrog scheme has the 3-metric and
extrinsic curvature variables offset by 1/2 time slice, so that
although the main time derivative terms are properly centered, a
number of important terms in the evolution equations require
extrapolations or averaging in time.  The details of these methods
have been published elsewhere [see e.g.Ref.~\cite{Bernstein89}], 
and so we will not present them here
(however see Section~\ref{V} where we apply these methods to a
simplified model problem).

\subsubsection{The ``H'' code}
\label{thehcode}

The second code (``H'') is based on the work of Bona and
Mass\'o\cite{Bona92} that casts the Einstein equations in an
explicitly first order, flux conservative, hyperbolic form.  In this
paper we present the first results of this new formulation to
gravitational wave spacetimes.

The general metric is also of the form (\ref{4metric}) and spacetime
coordinates are chosen such that the shift vector vanishes.  It was
shown in Ref.~\cite{Bona92} that if one restricts the lapse to the
harmonic slicing~(\ref{harmonic}), one can write the Einstein
evolution equations as a hyperbolic first order system of balance laws
that in vacuum takes the form:
\begin{eqnarray}
\partial_t g^{ij} &  = & Q^{ij} \label{harm1}\\
\partial_t\left[\frac{\sqrt{g}}{\alpha} Q^{ij}\right]
& - & \partial_k\left[\alpha\sqrt{g}\left(D^{kij} + g^{ki} g^j
      +g^{kj} g^i \right)\right]  \qquad \qquad \qquad \nonumber \\
      \qquad
      & = & \frac{\sqrt{g}}{\alpha} Q^{ik} Q^j_k -2\alpha\sqrt{g}
      \left[g^{ikl} g^j_{kl} + L^iL^j 
      - g^ig^j \right]~, \label{harm2} \\
\partial_t\left[D^{ij}_k\right] & - & \partial_k\left[ Q^{ij}\right] =
0~, \label{harm3} \\
\partial_tg^i & = &   Q^k_k L^i -2Q^i_jL^j +g^i_{jk} Q^{jk}~.
\label{harm4}
\end{eqnarray}
The $Q^{ij}$ quantities are proportional to the extrinsic
curvature. Note that all the sources (on the RHS) account for the
nonlinear terms and that the three-dimensional ricci does not appear
as it has been split into its transport part and its nonlinear source.
The connection coefficients $g^i_{jk}$ are constructed from the
first derivatives of the metric
\begin{equation}
D_k^{ij} = \partial_k g^{ij}~.
\label{drivare}
\end{equation}
These derivatives are evolved using Eq.~(\ref{harm3}).
Eq.~(\ref{drivare}) is only used in the initial slice. Similarly, the
derivatives of the lapse are used on the initial slice to construct
$L^i=\partial^i\ln\alpha$ and to derive the initial values of the momentum
constraint related variables:
\begin{equation}
g^i = \frac{1}{2}g_{jk} D^{ijk} - D_k^{ki} - L^i~.
\label{gamom}
\end{equation}
These variables are evolved using Eq.~(\ref{harm4})
while Eq.~(\ref{gamom}) is used to compute the $L^i$ during the evolution. 

At present this code is restricted to use the harmonic lapse
condition with a vanishing shift, although recent work~\cite{Bona94b}
shows that the same first order, flux conservative, hyperbolic form
can be maintained with a wider class of slicing conditions.  Results
from a code developed with this more recent formulation of the
equations will be presented elsewhere.

Standard operator splitting techniques allows for the principal part
of the system to be treated as a flux conservative first order system.
This kind of system is well known in Computational Fluid Dynamics
(CFD), where a wide choice of modern and standard numerical methods
have been developed.  In this case a flux conserving MacCormack method
is used for the principal part of the evolution system.  Note that
this is a {\em true} MacCormack method, developed for truly first
order systems of equations, and not the ``MacCormack-like''
predictor-corrector method used in the ``G'' code (again see 
Section~\ref{V} where we apply these methods to a
simplified model problem).

\subsection{The Linearized 3D Codes}
\label{linearizedcodes}

The discussion above was centered on the two codes we have developed
to solve the fully nonlinear Einstein equations.  In order to help
sort out linear from nonlinear effects and physical from numerical
effects, we have also developed linearized versions of both the ``G''
and ``H'' codes.  Both codes have been written in such a way that
subroutine calls can be made to solve either the full Einstein
equations or the linearized versions.  In this way all numerical
algorithms not associated with the expressions themselves are
identical and we can be sure that effects we see are related only to
the linearization process, and not to slight differences in coding or
numerical techniques that might otherwise arise if different codes
were developed. 

The general linearized version of the ADM Equations (\ref{dtgij}) and
(\ref{dtkij}) are long and unwielding to write out explicitly. The
task is simpler for the harmonic formulation, as it amounts to
linearizing the principal part and setting all the nonlinear sources
on the RHS of the equations (\ref{harm1})--(\ref{harm4}) to zero. In any case,
a simplified set of linearized ADM equations results when we set
$\beta^i=0$ and $\alpha=constant$ to second perturbative order.  We
will present these equations here to provide a framework for obtaining
analytic solutions to the Einstein equations at first perturbative
order for weak waves. However, we stress that it is the general form
of the linearized equations that we solve numerically, and not the
specialized equations presented below.

The perturbation expansion can be written in the form
\begin{eqnarray}
g_{ij} &=& g_{ij}^{(0)} + \epsilon g_{ij}^{(1)}~, \\
K_{ij} &=& K_{ij}^{(0)} + \epsilon K_{ij}^{(1)}~, 
\end{eqnarray}
where $\epsilon \ll 1$ is the smallness parameter and the superscripts
(0) and (1) refer to the $0th$ and $1st$ order solutions.  Assuming a
Minkowski background spacetime such that
\begin{eqnarray}
g_{ij}^{(0)} &=& \mbox{diag}~\left[1,~1,~1\right], \\
K_{ij}^{(0)} &=& 0~,
\end{eqnarray}
and $\alpha=1$, the $0th$ order equations are satisfied trivially and
the resulting linearized ADM equations become
\begin{eqnarray}
\partial_t g_{ij}^{(1)} &=& -2\alpha K_{ij}^{(1)}~, \label{linear1} \\
\partial_t K_{ij}^{(1)} &=&   \alpha R_{ij}^{(1)}~. \label{linear2}
\end{eqnarray}
If we make the further assumption of a diagonal 3-metric which is a
function only of the single coordinate $z$, the nonvanishing
components of the Ricci curvature tensor are
\begin{eqnarray}
R_{xx}^{(1)} &=& -\frac{1}{2} g_{xx,zz}^{(1)}~, \\
R_{yy}^{(1)} &=& -\frac{1}{2} g_{yy,zz}^{(1)}~, \\
R_{zz}^{(1)} &=& -\frac{1}{2} \left(g_{xx,zz}^{(1)} + 
                                    g_{yy,zz}^{(1)}\right)~.
\end{eqnarray}
Eqs.~(\ref{linear1}) and (\ref{linear2}) then reduce to three
equations for the diagonal metric components
\begin{eqnarray}
\partial_t^2 g_{xx}^{(1)} &=& 
             \alpha^2 g_{xx,zz}^{(1)}~, \label{lin1pde1} \\
\partial_t^2 g_{yy}^{(1)} &=& 
             \alpha^2 g_{yy,zz}^{(1)}~, \label{lin1pde2} \\
\partial_t^2 g_{zz}^{(1)} &=& 
             \alpha^2 \left(g_{xx,zz}^{(1)}
                           +g_{yy,zz}^{(1)}\right)~.
                                            \label{lin1pde3}
\end{eqnarray}
The Hamiltonian constraint (\ref{ham}) reduces to
\begin{equation}
R^{(1)} = -\left(g_{xx,zz}^{(1)} 
               + g_{yy,zz}^{(1)}\right) = 0~. \label{linham}
\end{equation}
Analytic solutions to equations (\ref{lin1pde1}) --- (\ref{linham})
are discussed in Section~\ref{linearized}.

\section{Code Test 1 - Single Wave Packet}
\label{III}

In this section we present a set of code tests involving the
propagation of plane wave packets traveling in one dimension.  We
evolve these plane wave packets with our full 3D codes to test wave
propagation in all three orthogonal directions ($x$, $y$ and $z$)
independently and to look for any asymmetries in the evolution for
debugging purposes.  These results can then be compared with the
propagation of waves along some arbitrary oblique angle that is not
parallel to any coordinate axis, which tests the accuracy of resolving
arbitrary waves on our rectangular grid.  Since for such waves we can
use fewer grid zones in the transverse directions than in the
longitudinal directions, this allows us to perform tests without
severely being constrained by available computer memory as in the full
3D case.
We have checked in all cases we have studied that for very low
amplitudes, the evolutions obtained by the full 3D non-linear codes
are indistinguishable from those obtained by the linearized codes
described in Section~\ref{formalisms} above.

\subsubsection{linearized solution}
\label{linearized}

A solution to the perturbation evolution equations~(\ref{lin1pde1})
--- (\ref{lin1pde3}) that is consistent with the Hamiltonian
constraint (\ref{linham}) can be given by
\begin{equation}
ds^2 = -dt^2 + (1+f(t,z))dx^2 + (1-f(t,z))dy^2 + dz^2~,
\label{linmetric}
\end{equation}
with $f(t,z)$ satisfying the linear wave equation
\begin{equation}
\partial_t^2 f(t,z) = \partial_z^2 f(t,z)
\label{linwave}
\end{equation}
for linearized plane waves propagating in the $z$
direction~\cite{Misner73}. Setting $g_{xx}^{(1)} = -
g_{yy}^{(1)}$ gives the transverse-traceless (TT) gauge in which
the wave amplitudes are purely spatial, traceless and transverse to
the propagation direction.  The metric (\ref{linmetric}) describes
gravitational waves with a single mode of polarization $e_+$.

We will study the solutions of a gaussian shaped wave packet with
\begin{eqnarray}
f(t,z) &=& \left[
                  A_{R}  e^{-(2\pi(t-z-a)/\sigma)^2} 
                  \cos\left(\frac{2\pi}{\lambda}(z-t)\right) \right.\nonumber \\
       &&\left.\qquad   + A_{L}  e^{-(2\pi(t+z-a)/\sigma)^2} 
                    \cos\left(\frac{2\pi}{\lambda}(z+t)\right)
         \right]~.
\label{ffunction}
\end{eqnarray}
The parameters $A_{R}$ and $A_{L}$ represent the amplitudes of waves
traveling to the right and left, respectively, with a gaussian shape
of width $\sigma$ and centered at $z=\pm a$ at $t=0$. 
$\lambda$ is the wavelength of the gaussian modulated
oscillations. If $\sigma \gg \lambda$, equation~(\ref{ffunction})
represents a pure sinusoidal mode and for $\sigma \ll \lambda$, a pure
gaussian packet.  By changing the metric functions appropriately, 
it can just as easily describe a wave traveling along the $x$- or
$y$-axes, or be generalized to a wave traveling in some arbitrary direction.

We note that the harmonic slicing condition (\ref{harmonic}) is
consistent with geodesic slicing ($\alpha=1$) to first order as long
as the traceless gauge ($K=0$) is maintained. Hence the linearized solutions
presented above apply to the hyperbolic formulation with no modifications.

\subsubsection{Convergence studies}
\label{convergence}

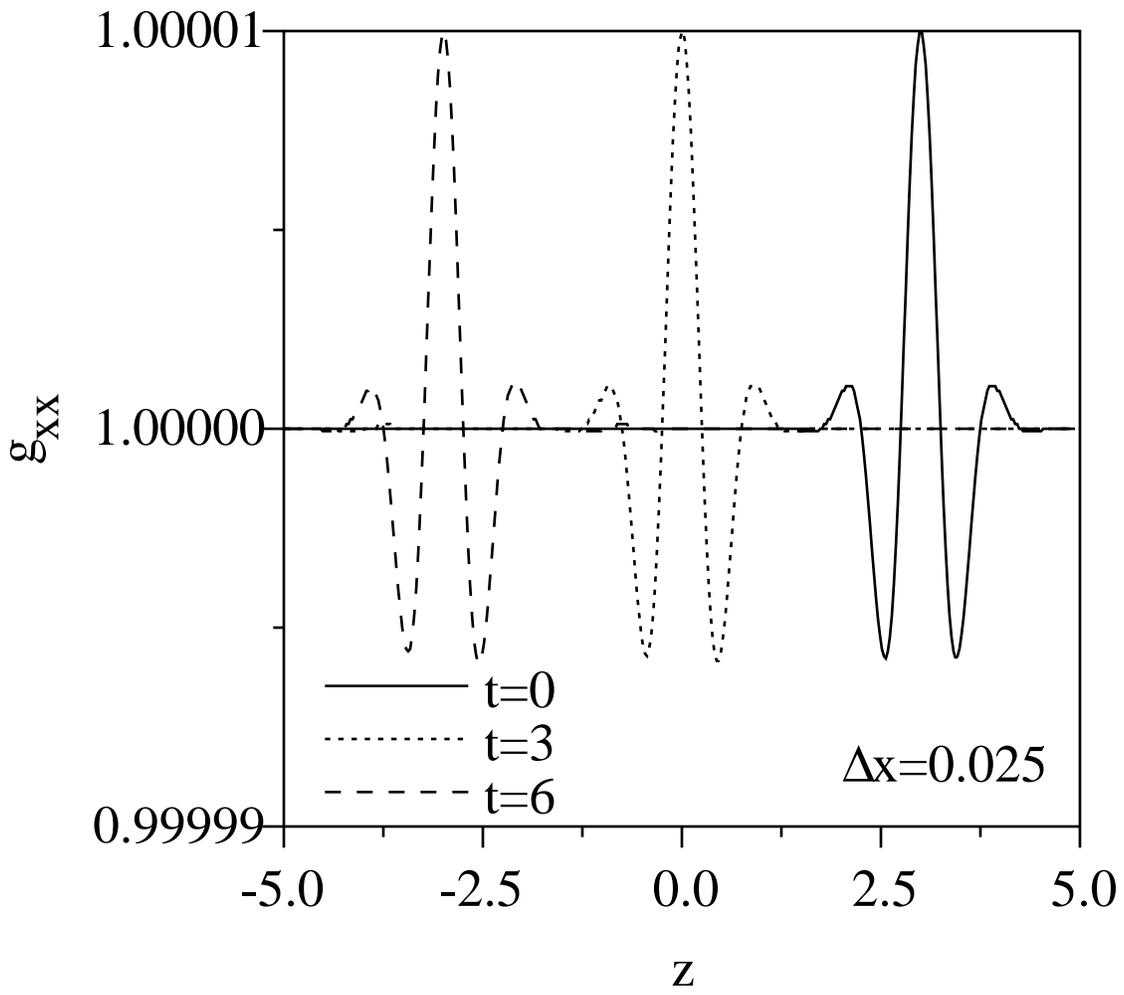
\begin{figure}
\caption{The evolution of the metric function $g_{xx}$ is shown for a
plane wave with shape parameters $\sigma=2.0$, $\lambda=1.0$,
$A_{L}=0.00001$, $A_{R}=0$, and $a=3$.  The wave is shown at
times; t=0, t=3, and t=6.  
This wave was evolved with 40 points per wavelength}
\label{fig:nice_wave}
\end{figure}

In Fig.~\ref{fig:nice_wave} we show the evolution of the plane
symmetric waves defined by (\ref{linmetric}) and (\ref{ffunction})
with shape parameters $\sigma=2.0$, $\lambda=1.0$, $A_{L}=0.00001$,
$A_{R}=0$ and $a=3$.  This run is typical of the resolution and time
scales for most of our evolutions.  The wave is shown at $t=0$, $t=3$,
and $t=6$.  The evolution is with $\Delta x=\Delta y=\Delta z=0.025$.

In Fig.~\ref{fig:planewave} we evolve the initial data above, but with
$a=0$ and periodic boundary conditions.
This allows the waves to continue to propagate through the
computational domain, allowing us to evolve the wave for much
longer times, without increasing the grid size. 
The waves are shown at three different times
$t=0$, 10 and 20. Since the wave propagation speed is unity and the
outer grid boundaries are set at $z=\pm 5$, the displayed profiles
correspond to the wave positioned at the grid center. At $t=20$, the
wave has propagated across the extent of the entire grid twice.  Data
for the same sequence of times are presented for three different
spatial resolutions with grid spacing $\Delta x = 0.1$, 0.05 and 0.025
for both the ``G'' and ``H'' codes.  The ``G'' code evolutions are
performed with the standard leapfrog scheme with half step
extrapolation. A full MacCormack scheme is used in the ``H'' code.

At the coarser resolutions, the waves disperse due to numerical
discretization effects. These effects are more evident in the ``H''
code evolutions of Fig.~\ref{fig:planewave}a.  At higher resolutions,
the two codes yield comparable results that reproduce accurately the
solution~(\ref{ffunction}), which is represented by the initial data
at $t=0$.  

In Fig.~\ref{fig:errorpw} we plot the RMS error, where the error is
defined as
\begin{equation}
E = 
    \left| \frac {g_{xx}^{(a)} -
g_{xx}^{(n)}} {g_{xx}^{(a)}} \right|~,
\label{error}
\end{equation}
as a function of the grid resolution $\Delta x$. Here
$g_{xx}^{(a)}$
is the linear analytic solution (\ref{ffunction})
and $g_{xx}^{(n)}$ is the numerical solution from the nonlinear codes.
[As the amplitude of the wave is low, the analytic solution to the
linearized equations Eq.~\ref{ffunction} is basically the same as the
exact nonlinear solution, and $E$ in Eq.~(\ref{error}) represents the
error in this sense.]
The boxes (circles) are the ``G'' (``H'')
code results.  We find the error scales as $E\sim \Delta x^\alpha$
with $\alpha \sim 2$ as expected for fully second order methods.  In
all our simulations, the timesteps have been chosen to be proportional
to the grid spacing $\Delta t = C \Delta x$, with $C<1/\sqrt{3}$ to
satisfy the 3D Courant stability condition.  We use $C=0.2$ for both
codes in the calculations presented in this section.  
We find that in order to keep errors below
$E < 10^{-4}$ at $t=10$, it is necessary to resolve a wavelength
with 20 grid points with the
``G'' code  and 40 using the ``H'' code.
For waves traveling along the diagonal, we find the resolution needs
to be increased by approximately $\sqrt{2}$ to get the same error as when
the wave is traveling along an axis, as expected.

By looking at the solutions in Fourier space, we can see numerical 
effects not clearly evident in Fig.~\ref{fig:planewave}.  In 
Fig.~\ref{fig:fftpw} we plot the Fourier transform of $g_{xx}-1$ at 
three different times for the intermediate resolution case with 
$\Delta x = 0.05$.  The wavelength $\lambda = 2\pi / k = 1$, 
corresponding to the dominant mode, is resolved with $\sim 20$ grid 
cells at this resolution.  We find, in general, that amplitude errors 
due to numerical dissipation dominate over phase errors for typical 
resolutions, and that the MacCormack method used in the 
``H'' code is significantly more dissipative and dispersive than the 
leapfrog method of the ``G'' code.  Again, we stress that 
this is what we expect from the mathematical properties of 
their respective finite differencing operators.

As another test of the code, we monitor the Riemann curvature
invariants \cite{Kramer80}.
It is known \cite{Kramer80} that spacetimes containing only plane-fronted
gravitational waves with parallel rays ($pp$ waves) are of the Petrov
classification type $N$ and have vanishing invariants.
We therefore expect
(at least to linear order at which the metric (\ref{linmetric}) satisfies
Einstein's equations) both curvature invariants $I$ and $J$ to vanish.
The invariant $I$ 
is plotted in Fig.~\ref{fig:curv_pw} at three
different resolutions to see that it is indeed converging to zero.

\section{Code Test 2 - Colliding Waves}
\label{IV}

The propagation of plane symmetric waves discussed in the previous
section allows many aspects of the codes to be tested, including the
dispersive and dissipative nature of the various numerical schemes.
Here we consider the collision of two identical plane wave packets.
In such cases one expects to find nonlinear effects, even for
vanishingly small amplitude wave packets.  In fact, it is known that
when two plane symmetric waves collide when traveling through an
otherwise flat background, a curvature singularity is generated in the
region where the waves cross due to the focusing effect of the waves
~\cite{Khan71}.  Such a singularity gets generated even for
arbitrarily weak waves, only the singularity will emerge at a later
time.

In Figs.~\ref{fig:colliding}a--e we show an
evolutionary sequence of a wave packet collision at the four times
$t=0$, 3, 6 and 9 for moderately resolved grids with $\Delta x = 0.05$
for Figs.a-d and $\Delta x = 0.025$ for Fig.6e.
The initial data is of the form of Eq.~(\ref{ffunction}) with the same
parameters as the single wave packets in the previous section except
now $a=3$, and $A_{L}=A_{R}=0.025$ so that the data set consists 
of two wave packets centered
at $z= \pm 3$.  
First the two waves approach each other from their initial
configurations at $t=0$, collide at $t=3$, and propagate to their
original centered locations at $t=6$.  

Notice that in the ``G'' code
after the collision there is a remnant left behind the waves.  This
remnant, shown clearly in Fig.~\ref{fig:colliding}a, grows in time.  
For waves with
smaller amplitude, this remnant is smaller initially, but grows to a
large value at late times.

To test if the remnant in Fig.~\ref{fig:colliding}a is a nonlinear
effect, in particular, if it is related to the singularity due to the
focusing effect, we evolved the same initial data set using the
{\it linearized} evolution equations.  With linear evolution, no focusing is
possible.  The results are displayed in Fig.~\ref{fig:colliding}b.
There is no remnant in the solutions for colliding linear plane waves.
In view of this, one might be tempted to conclude that the remnant in
case(a) is due to nonlinear physics. In fact, we will show this is not
the case.

In Fig.~\ref{fig:colliding}c we show results from a ``G'' code
simulation using the same initial data and resolution but with the
MacCormack scheme. The remnant is greatly reduced.  We also show in
Fig.~\ref{fig:colliding}d the equivalent simulation performed with the
``H'' code.  Here we see similar behavior as the waves approach and
collide. However, after the collision we see that the remnant is
nearly nonexistent, and it does not grow appreciably over the time
scale of the run.  Clearly the different numerical methods produce
different results in the evolution.  Finally, in
Fig.~\ref{fig:colliding}e we show the same simulation with the fully
nonlinear ``G'' code as before in Fig.~\ref{fig:colliding}a, but now
with twice the resolution.  In this case all other features are quite
similar, but the remnant is now reduced significantly in amplitude.
If we again double the resolution we will see the remnant reduced even
further.  We conclude that the remnant observed in
Fig.~\ref{fig:colliding}a is a numerical artifact dependent on the
numerical method and grid resolution. 

So this remnant is {\it not} related to the singularity caused by the
focusing effect.  On the other hand, we know that there must be a
singularity at a later time; how does it manifest itself?  We note that
the weaker the amplitude of the wave, the later in time the
singularity will form. Based on the colliding packet study in
Refs.~\cite{Yurtsever88a} and \cite{Yurtsever88b},
 we expect that the singularity will develop at a
time
\begin{equation}
t \sim \frac {\lambda ^{2}} {(2\pi)^{2} \sigma A^{2}}
\label{sing}
\end{equation}
after the collision.  Where $\lambda$ is the characteristic
wavelength, $\sigma$ is the characteristic width of the packet, and $A$ is the
characteristic amplitude of the packet.  For the case here, with
$\lambda \sim 1$, $\sigma \sim 1$, and $A \sim 10^{-2}$, we expect the
singularity to appear at $t \sim 250$, which is far beyond any
evolutions shown here.  In fact it is well beyond any time we can
accurately evolve to with our present computer resources.  It is
tempting to make the singularity appear earlier by increasing the
amplitude of the waves, so that the onset of the singularity can be
studied.  We have resisted the temptation to do this here, mainly
because such a study is out of the scope of this paper.  Another
reason for not including this study in the paper is that, for a larger
amplitude wave, one has to solve the initial constraints to higher
order.  With the planar symmetry, the non-linear effect of
the wave will introduce a long length scale variation in the metric,
which causes a coordinate singularity at some spatial
location on the initial slice, and hence requires special treatment.

\section{A Model Nonlinear Problem}
\label{V}

To investigate the cause of the ``remnant'' in the nonlinear
evolutions, we have developed a simplified model problem containing a
single scalar field that exhibits similar behavior as the fully
nonlinear Einstein equations.

We arrive at this nonlinear model by starting with the metric
(\ref{linmetric}) used in the previous studies.  However, now we keep
the nonlinear terms in the ADM evolution equations (\ref{dtgij}) and
(\ref{dtkij}).  These lead to the evolution equation for $f(t,z)$
\begin{eqnarray}
\partial_t f &=& \Pi~, \label{model1}\\
\partial_t \Pi &=& f_{,zz} + 
                  \frac{\Pi^2 - (f_{,z})^2}{1-f^2}~. \label{model2}
\end{eqnarray}
Together, Eqs. (\ref{model1}) and (\ref{model2}) become
\begin{equation}
f_{,tt} - f_{,zz} = \frac{(f_{,t})^2 - (f_{,z})^2}{1-f^2}~.
\label{model4}
\end{equation}
When the order $f^2$ term on the R.H.S. is negligible,
Eq.~(\ref{model4}) reduces to the standard wave equation
Eq.~(\ref{linwave}).  Our aim here is to investigate the relation between
this $f^2$ term and the numerical schemes used for the evolution.
 We note that the solution of Eq.~(\ref{model4})
does not generate a solution of the Einstein equations as the
resulting metric does not satisfy the constraint equations.

We have investigated this model equation using several different
finite difference methods that closely parallel those used in the
``G'' and ``H'' codes.  Here we will present results for the two
methods used in the ``G'' code; staggered leapfrog with 1/2 time step
extrapolation and a MacCormack-like predictor corrector scheme with no
extrapolation.

For a full understanding of the effect, we give the complete
discretized equations, first in the leapfrog scheme
\begin{eqnarray}
f^{n+1}_{j}   &=& f^{n}_{j} + \Pi^{n+1/2}_{j} \Delta t \\
\Pi^{n+3/2}_{j} &=& \Pi^{n+1/2}_{j} + \left[(f^{n+1}_{j})_{,zz} +
        \frac{(\Pi^{n+1}_{j})^2 - ((f^{n+1}_{j})_{,z})^2}
        {1-(f^{n+1}_{j})^2}\right] \Delta t~,
\end{eqnarray}
where the superscript $n$ denotes the time level, and subscript $j$
tracks the spatial position. $f$ and its time derivative $\Pi$ are
staggered by a half step in time with respect to each other.  Note
that in updating the auxiliary variable $\Pi$ from time $n+1/2$ to
$n+3/2$, we need $\Pi^{n+1}$, but in the standard leapfrog scheme this
auxiliary variable only exists on the half time steps.  We approximate
this value by extrapolating data from the previous two time steps
\begin{equation}
        \Pi^{n+1} = \frac{3}{2} \Pi^{n+1/2} - \frac{1}{2}
\Pi^{n-1/2}~.
\label{extrap}
\end{equation}

In the MacCormack scheme, we first solve the predictor step for the
intermediate variables $\tilde f$ and $\tilde \Pi$
\begin{eqnarray}
\tilde f^{n+1}_{j} &=& f^{n}_{j} + \Pi^{n}_{j} \Delta t~, \\
\tilde \Pi^{n+1}_{j} &=& \Pi^{n}_{j} + \left[(f^{n}_{j})_{,zz} +
        \frac{(\Pi^{n}_{j})^2 -((f^{n}_{j})_{,z})^2}
        {1-(f^{n}_{j})^2}\right] \Delta t~,
\end{eqnarray}
followed by the corrector step
\begin{eqnarray}
f^{n+1}_{j} &=& \frac {1}{2} \left[\tilde f^{n+1}_{j} + f^{n}_{j} +
        \tilde \Pi^{n+1}_{j} \Delta t\right]~, \\
\Pi^{n+1}_{j} &=& \frac12\left[\tilde \Pi^{n+1}_{j} + \Pi^{n}_{j} +
        \left((\tilde f^{n+1}_j)_{,zz} +
        \frac{(\tilde \Pi^{n+1}_{j})^2 - ((\tilde f^{n+1}_{j})_{,z})^2}
        {1-(\tilde f^{n+1}_{j})^2}\right) \Delta t\right]~.
\end{eqnarray}
In this scheme, all variables are centered at the same time step at the
completion of both predictor-corrector updates. 

Results for the collision of two wave packets are shown 
in Fig.~\ref{fig:model}a.  The
initial data is given by Eq.~(\ref{ffunction}) with parameters
$\sigma = 1$, $A_{L} = A_{R} = 0.1$, and $a = 3$. 
We also set $\lambda \rightarrow
\infty$ so that the initial data is a pure gaussian wave packet
without sinusoidal oscillations. All calculations presented here were
run at the same grid resolution of $\Delta x = 0.05$.  Although the
leapfrog and MacCormack schemes both perform well on the standard
linear wave equation, they behave quite differently on this nonlinear
test problem.  In Fig.~\ref{fig:model}b  we zoom in
on the flat central portions of the afterwake.  Although only the
results from the leapfrog evolution are shown, we see a similar
drift with the MacCormack-like evolution, although the drift is orders
of magnitude smaller.

To understand these drifts, we note that under the approximation $f
\ll 1$ and $f_{,z}=0$, which are clearly appropriate in the region of
the drift, Eq.~(\ref{model4}) reduces to
\begin{equation}
f_{,tt}  = f_{,t}^2 
\label{model5}
\end{equation}
which has a solution
\begin{equation}
f(t) = - \ln(c_1 t + c_2)
\label{model6}
\end{equation}
where $c_1$ and $c_2$ are arbitrary constants.  To verify that the
drifts are really of this form, we look at the origin after the waves
pass through each other and then plot the quantity $e^{-f} - 1$ versus
time.  The results for the two different numerical methods are shown
in Fig.~\ref{fig:modeldrift} for $\Delta x = 0.05$, and
indeed, we see straight lines.  The constants $c_{1}$ and $c_{2}$ in
Eq.~(\ref{model6}) can be read out from the slopes and intercepts of
these curves.  For this case, we find the MacCormack
scheme has a much smaller drift rate with $c_1=1.4$x$10^{-7}$ and
$c_2=1.0$, as compared to the leapfrog scheme with
$c_1=-9.8$x$10^{-5}$ and $c_2=1.0$.  Just as in the full ``G'' code,
the remnant amplitude gets smaller as one goes to higher and higher
resolution.  We find that the drifting solution converges away
with rates $3.86$ and $3.49$ for the leapfrog and MacCormack-like methods
respectively. Here we are just using the three values of $c_{1}$ at
different resolutions to calculate the convergence rate $\alpha $
\begin{equation}
\alpha = \frac {c_1(\Delta x=0.05) - c_1(\Delta x=0.025)}
{c_1(\Delta x=0.025) - c_1(\Delta x=0.0125)} .
\end{equation}
(We note that this unstable mode can also be excited by a single wave
packet and appears in the tail after the wave passes some region.)

The drifts shown in this section for the evolution of
Fig.~\ref{fig:modeldrift}, and in the previous section for the
evolution of the full Einstein equations are now readily understood:  The
nonlinear evolution equations contain unstable modes.  We note that
this is not in contradiction to the expectation that the Einstein
equations are stable for weak waves (weak perturbations of the flat
spacetime).  It is the constraint equations that rule out these
unstable modes.  In our free evolution code the constraint equations
are not enforced.  This allows the unstable modes to develop after they
are excited by the numerical errors in the evolution.  Exactly which
mode will be excited most and the amount it is excited depends on the
details of the numerical scheme.  Here we see that the leapfrog scheme
as given by (\ref{model1}) and (\ref{model2}) is more prone to the
excitation of the unstable modes of the form (\ref{model6}).  This
is because the extrapolation (\ref{extrap}) leads to inaccuracies that
ruin the exact cancellations on the RHS of (\ref{model4}) in the
trailing edge of the wave.  We have further analyzed this point by (i)
studying the unstable mode given by (\ref{model6}) for the case of
a single wave packet, in which the same phenomena occurs; (ii) using
a different extrapolation scheme in place of that given by
(\ref{extrap}), e.g. one based on a second-order Taylor expansion; and (iii) by
reducing the Courant factor by a factor of 10.  We find that
increasing the accuracy of the extrapolation in Eq.~(\ref{extrap}) 
leads to slightly better results as far as the unstable drift
is concerned, but none of the methods we tried compare favorably
to the predictor-corrector schemes which require no
extrapolation.

To confirm that this is the same phenomenon as we observed in the 
Einstein equations, we have verified that the drift in the wave
remnants follows the form (\ref{model6}).
For similar grid parameters we find similar values for the coefficients:
$c_1=-1.6$x$10^{-4}$ and $c_2=1.0$ for the leapfrog
method and $c_1=7.9$x$10^{-7}$ and $c_2=1.0$ for MacCormack-like
method.  Again, the drifting solution is orders of magnitude smaller
for the MacCormack-like method.

\section{Code Test 3 - Pure Quadrupole waves}
\label{VI}

The third test problem on the construction of general relativistic
spacetimes we discuss is the quadrupole
waves~\cite{Eppley79,Teukolsky82} with an
imploding---exploding nature.  We use the quadrupole waves to test the
3D propagation of low amplitude waves in our 3D Cartesian codes.  As
these solutions represent quadrupole waves, they provide standards
against which we can compare the codes' ability to evolve waves which
do not conform to the rectangular geometry of Cartesian grids.  In the
following two subsections, we study these waves first in linear
settings and then with full nonlinearity.

\subsection{Quadrupole waves satisfying the IVP to linear order}

Linearized quadrupole waves (Teukolsky waves) have been given for both
even and odd 
parity solutions and the independent azimuthal modes in
Ref.~\cite{Teukolsky82} . Due to the length of these expressions, we
do not write out the solutions here.  The axisymmetric version of
these solutions has been used as a testbed for a number of
axisymmetric evolution codes (see, for example, Ref.~\cite{Evans86}).

In our first set of numerical tests, the initial data is taken to be 
essentially the form given by~\cite{Teukolsky82}, but modified to be time 
symmetric and contain an ingoing and outgoing wave in such a 
combination as to make them regular everywhere in spacetime 
~\cite{Eppley79}.  We note that as small amplitude waves on the 
Minkowski background, the constraint equations are trivially 
satisfied to first order, but violated to second order.
Quadrupole waves that satisfy 
the full constraint equations
will be studied in the next section.

We study the evolution of the waves using both the ``G'' and
``H'' codes.  The G code is run with geodesic slicing, and the H
code with harmonic slicing.  We first look at runs with even parity
waves having an amplitude of $10^{-5}$ and azimuthal mode number
$m=0$.  Here the amplitude is the amplitude given by the Eppley
packet~\cite{Eppley79} which corresponds to a perturbation in the metric
function $g_{xx}$ of about $0.025\%$.  
For such low amplitude waves, the difference coming from 
nonlinearities in the Einstein equations is negligible.
Initially the wave is at
the coordinate center and expands outward as time increases.
Fig.~\ref{fig:quadrupole_small}a plots $g_{xx}$ at various times
obtained by the ``G'' code and in Fig.~\ref{fig:quadrupole_small}b we
blow up the region near the axis to show the wave in the metric
function that rapidly falls off. 

By $t=5$, $g_{xx}$ evolves to become nearly unity everywhere.
Comparing the profile at $t=5$ to the
linearized solution in~\cite{Teukolsky82}, we find that the error
in $g_{xx}$ is about $1.4$x$10^{-6}$.
Fig.~\ref{fig:quadrupole_small}c shows the evolution with
the ``H'' code.  We see that the results of the two codes are similar.
The error in the ``H'' code at $t=5$ is about $1.1\times 10^{-7}$.
If we require that the error remains $< 10^{-6}$, at $t=5$ we see
that we need $40$ points/$\lambda$ for the ``G'' code and
$10$points/$\lambda$ for the ``H''.  The dispersive nature of the
``H'' code is probably biasing this result by allowing the waves to
disperse out faster.

In Fig.~\ref{fig:errortw} we plot the error, as defined by
Eq. (\ref{error}), in $g_{xx}$ as a function of grid resolution
at time $t=1$.
Again we observe a convergence rate with an exponent of nearly two.
We have also compared other metric components and various
components of the Riemann tensor, and they all showed results
agreeing to high accuracy with the linear analytic solution.

Next we study a case of higher amplitude perturbations with $A =
10^{-4}$ and $g_{xx} - 1 \sim 10^{-3}$.  The evolution using the
``G'' code with geodesic slicing, and resolution $\Delta x=\Delta
y=\Delta z=0.1$ is shown for late times in Fig.~\ref{fig:quadrupole}a.
The feature 
to note is that $g_{xx}$ develops a dip at the origin.  To
distinguish if the dipping is due to numerical or physical nonlinear
effects coming from the increased amplitude, we ran the same initial
data with the linear evolution equation option of the code.  The
result is shown in Fig.~\ref{fig:quadrupole}b.  No dipping is present
whatsoever.  This confirms that it {\it is} a nonlinear effect.  As we
pointed out earlier, there can be three types of non-linear effects:
(i) numerical errors coupled with nonlinearity, (ii) coordinate
effects due to nonlinearity, or (iii) nonlinear physics.  We expect
all three types to be present in the evolution.  The question is,
which one is most responsible for producing this dipping feature.

One might be tempted to identify this dip with the same spurious
drifting coming from the coupling of the finite differencing error and
the nonlinear term discussed in the previous section, namely effect
(i).  Both the drift in the previous section and the dip here are
secular evolutions in the region where the wave has passed.  However,
there is a major difference.  In this case, the dipping is not
converging away with higher resolution.  In Fig.~\ref{fig:quadrupole}c
we show the same quantities now evolved with $\Delta x=\Delta y=\Delta
z=0.05$ The dipping becomes slightly worse with the resolution
doubled.  We have carried this out at even higher resolutions with
runs up to $\Delta x=\Delta y=\Delta z=0.025 $.  We conclude that the
dipping is not due to finite differencing error.

At this point we want to investigate another possibility for the cause
of the dipping, which is not included in (i)-(iii) mentioned above.
We note that the initial data set that we have used
satisfies the initial constraint equations to first order only.
While we evolve the initial data with the full nonlinear evolution
equation, is it possible that there may be spurious effects due to this
contradiction that leads to the dipping?  This is the subject of the
next subsection.

\subsection{Quadrupole waves satisfying the IVP}

To generate a set of initial data which is similar to what is studied
in the previous subsection, we take the linear data as the metric
$\hat g_{xx}$ in the conformal space in the York
formalism~\cite{York79}.  As the linear data set is constructed to be
time symmetric with $\hat K_{ij}=0$, the initial momentum constraint
equations are trivially satisfied and it is straightforward to solve
the initial Hamiltonian constraint equation to determine the conformal
factor $\Psi$ needed for the physical space metric $g_{xx} =
\Psi^4 \hat g_{xx}$.  For the case where the amplitude is taken
to be $10^{-4}$ (the $\hat g_{xx}$ of which is given in
Fig.~\ref{fig:quadrupole}) the conformal factor is shown in
Fig.~\ref{fig:conf}a.  We note that $\Psi$ differs only slightly from
1, so that the initial data obtained through this procedure describes
basically the same spacetime as studied in the previous subsection,
except that now the initial data satisfy the constraint equation in
full, and can be regarded as representing a physical spacetime as
described by the Einstein equations up to the finite differencing
approximation.

The evolution of this initial data is shown in Fig.~\ref{fig:conf}b
again using the G code with geodesic slicing, and $\Delta x=\Delta y =
\Delta z =
0.05$.  We see that the time development is basically the same as in
that of Fig.~\ref{fig:quadrupole}a.  In particular, the dipping at the
origin at late times is not affected.

After verifying that the dipping is not due to numerical truncation
error (the effect doesn't decrease with resolution), and that it is
independent of 
whether the IVP is solved or not, there are two possibilities left.  
The dipping is either due to (ii)
nonlinear coordinate effects, or (iii) nonlinear physics, as
discussed and labeled in the introduction section.  To
distinguish which one is the main cause, we look at variables that are
representative of the actual geometry.  We studied various components
of the four dimensional Riemann tensor, e.g.,
$\Re_{\alpha\beta \gamma\delta}$, the Riemann invariants $I$ and $J$
\cite{Kramer80}, 
and the pseudo energy-momentum tensor
~\cite{Weinberg72}
\begin{equation}
\tau_{\mu\nu} = \frac{1}{8\pi G} \left(\Re_{\mu\nu} 
            -\frac{1}{2} g_{\mu\nu} \Re
            - \Re^{(1)}_{\mu\nu} 
            +\frac{1}{2}\eta_{\mu\nu} \Re^{(1)} \right)~,
\label{energy}
\end{equation}
where $\Re^{(1)}_{\mu\nu}$ is the part of the four dimensional Ricci
tensor that is linear in the deviation of the metric from flat
spacetime.  For simplicity, we assume $\Re_{\mu\nu}=0$ when evaluating
Eq.~(\ref{energy}) numerically.  We note that this $\tau_{\mu\nu} $ 
is meaningless if the
initial data satisfy the constraints only to the linear order.  For
this reason $\tau_{\mu\nu}$ is not used in the analysis of any of the
linearized initial data in the previous section.  

In Figs.~\ref{fig:slicing_tw}a-g we compare the ``G'' code geodesic slicing
evolution to the maximal slicing case.  Notice that (1) even while the
metric is dipping in the geodesic slicing case all the components of
the Riemann tensor studied, the Riemann invariants, and $\tau_{tt}$,
all remain small, and are consistent with returning to zero at late
times (see Figs.13b and 13c).  (2) In the maximal slicing case, there
is no dipping of 
metric components (Fig.13d).  (3) There is good agreement in the Riemann
tensor components, the Riemann invariants, and $\tau_{tt}$ between the
geodesic and the maximal slicing cases (Figs. 13e and 13f), although the
metric functions behave differently.  In Fig.~\ref{fig:slicing_tw}g,
we show the evolution of lapse in the maximal slicing case.  We see
that the lapse is very close to one throughout the
spacetime.  This means that in
terms of proper time evolved, the geodesic slicing case and the
maximal slicing case are not that different.

This strongly suggests that the dipping should be attributed to 
nonlinear coordinate effects.
The energy of the wave initially sitting at the origin sets the 
coordinate lines (which move normal to the slicing in the case of zero
shift) into free fall 
towards the origin.  As the wave moves outward, the geometry near the 
origin returns to being flat.  However, with geodesic slicing and no 
shift vector, there is nothing to stop the motion of coordinate lines.  
They keep drifting towards the center where the wave was, causing
the metric functions to dip there.  With maximal slicing, 
the motion of the coordinate lines is changed as the normal of the 
constant time surface changes with respect to the four geometry. This
is enough to stop the secular motion of the coordinate lines in this 
weak field case without having the lapse collapse in any significant 
manner.  This kind of gauge problem in evolving with geodesic slicing 
is well noted in the literature\cite{Smarr78b}).  From 
Eq. (\ref{dtkij}) we can compute the evolution of $K$ for geodesic 
slicing, which, using the hamiltonian constraint reduces to:
\begin{equation}
	\partial_{t} K =  K_{mn}K^{mn}\;.
	\label{geodesicK}
\end{equation}
The RHS of this equation is always non-negative.  Therefore, the 
convergence $K$ of the geodesics tends to increase without limit, 
resulting in a coordinate singularity on a free-fall time scale.  See 
Ref.~\cite{Smarr78b} for a full discussion.  Here we found that the 
dipping seen in Fig.~\ref{fig:slicing_tw}a is due solely to this 
effect.

By comparing the metric functions obtained from a linear evolution to
a nonlinear evolution, we can define a qualitative measure of the time
at which nonlinear coordinate effects become present.  We do this by
defining $t_{critical}$ as the time when the RMS relative difference
of the linear and nonlinear evolutions disagree by $10\%$.  Since this
RMS value is a global measure, we expect our results to depend on the
specific energy distribution of the wave model that we are evolving.
In Fig.~\ref{fig:t_crit}a we compare the critical time as a function
of the size of the initial metric function perturbation.  We see that
the critical times scales roughly as a power law.  The error bars in
the graph come from the fact that the data is only analyzed in time
intervals of $\Delta t=0.1$.

To determine the time at which nonlinear geometric effects occur, we
define a similar critical time, but now comparing the RMS relative
difference of the linear and nonlinear evolutions of the curvature
invariant $I$.  Again, we define $t_{critical}$ as the time at which
the two evolutions disagree by $10\%$.  The results are shown in
Fig.~\ref{fig:t_crit}b, again plotted against the size of the initial
metric function perturbation.  The critical time for nonlinear
geometric effects occurs at a later time than that of nonlinear
coordinate effects for the amplitudes considered here.

\section{Conclusion}
\label{VII}

In this first paper in the series, we studied various aspects of our
3D codes in evolving gravitational waves.  We show how the accuracy of
the evolution can be analyzed through various monitors built into the
codes.  This includes violation of the hamiltonian constraint, Fourier
spectrum analysis, as well as 
convergence tests.  These studies are not only crucial for our using
these codes in the future, but are also useful for other groups who
may want to build similar 3D codes.

We focused on the difficulties in evolving low 
to moderate amplitude gravitational
waves.  They have amplitudes low enough so that one has a good physical
understanding of the physics involved, but at the same time large
enough to enable non-linear effects to emerge.  We studied (i) the
coupling between numerical errors and nonlinearity, and (ii) coordinate
effects due to nonlinearity, with specific examples.  We discussed the
strategies used in identifying the cause of the non-linear effects.
In this process we emphasize the importance of the flexibility of being
able to use different numerical schemes, different choice of
coordinate conditions, different formulations of the Einstein
equations (G and H formulations), and different equations (linear
vs. nonlinear equations).  This flexibility, and the availability of
many ``monitoring devices'' in the codes, such as the scalar Riemann
invariants, pseudo energy-momentum tensor, and hamiltonian constraint,
have been 
crucial in our understanding of the nonlinear effects.

With these in hand, we are now proceeding to study the collision of 3D
wave packets (packets finite in size in all three spatial dimensions).
We consider this to be possibly just next in importance in
geometrodynamics to the collision of two black holes.  The results will
be reported in later papers in the series.

\acknowledgements 

We thank Steve Brandt for coding up the Riemann invariant
routines used in this paper. This work was supported by NSF grants 
PHY94-04788, 94-07882, and 96-00587.  The calculations were
performed at NCSA on the Thinking Machines CM-5 and at the Pittsburgh
Supercomputing Center on the Cray C-90.  WMS would like
to thank the 
support of the Institute of Mathematical Sciences of the Chinese
University of Hong Kong.  In the late stage of the preparation of this
paper, a paper by Shibata and Nakamura appeared (Ref.~\cite{Shibata95})
reporting on their numerical 
study of gravitational waves based on a code they developed independently.



\begin{figure}
\caption{The evolution of the metric function $g_{xx}$ is shown for a
plane wave with shape parameters $\sigma=2.0$, $\lambda=1.0$,
$A_{L}=0.00001$, $A_{R}$, and $a=0$.  Periodic boundary
conditions are applied to allow the wave to evolve for a long time.  
The evolutions (a)-(c) are done with the H
code, and the evolutions (d)-(f) are done with the G code.  Here we
see the effects of dispersion when insufficient resolution is used.}
\label{fig:planewave} 
\end{figure}

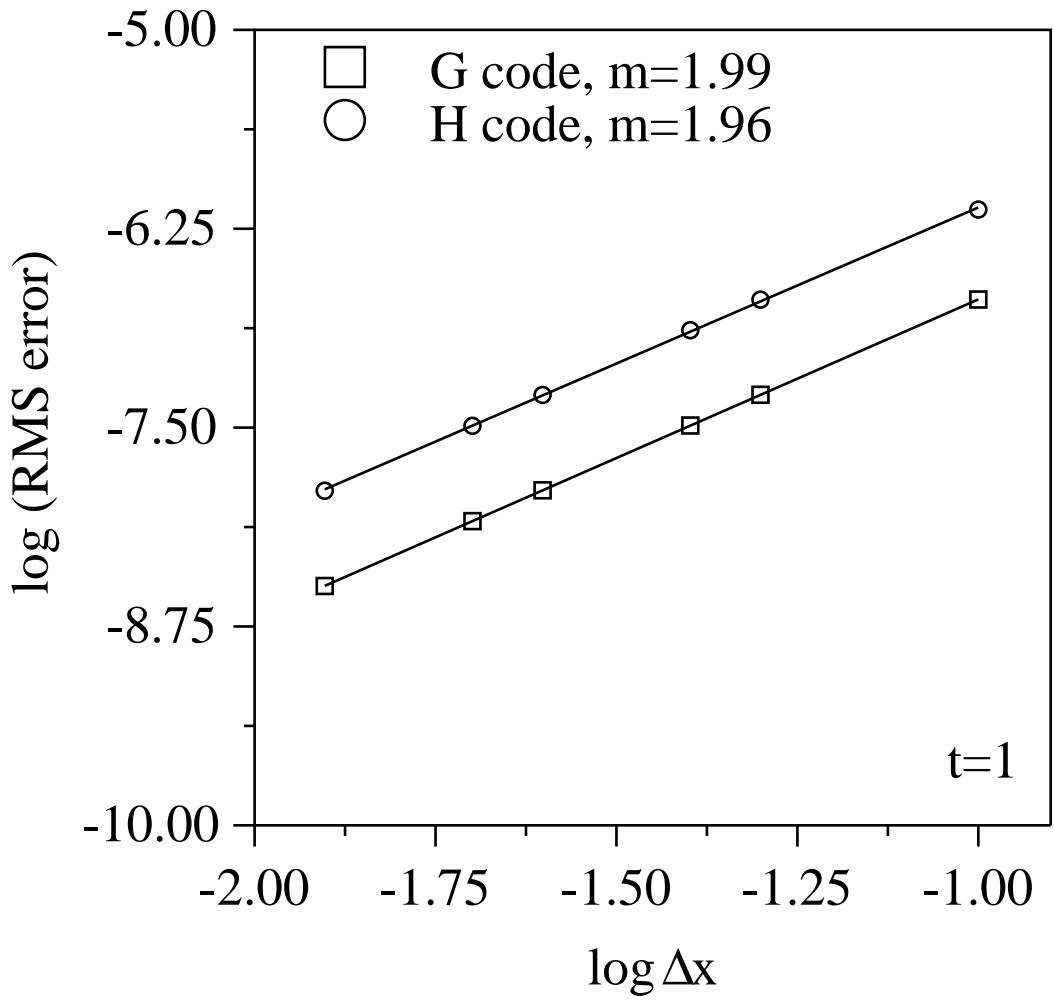
\begin{figure}
\caption{The log of the RMS error is plotted against the log of the 
resolution $\Delta x$, to test the convergence of the code.  Here the 
error $E$ is defined in the text with respect to the linear solution.  
Although we are evolving the solution with the full non-linear 
equations, with the small amplitudes used, we expect the wave to 
behave linearly.  Second order methods were applied throughout, so we 
expect the slope of this graph,m, to be 2.}
\label{fig:errorpw} 
\end{figure}

\begin{figure}
\caption{The real part of the Fourier transform of the metric 
functions plotted in Fig. 2b, and Fig. 2e are shown to compare the 
effects of dispersion and dissipation.  The H code is found to be more 
dissipative and dispersive than the G code.}
\label{fig:fftpw} 
\end{figure}

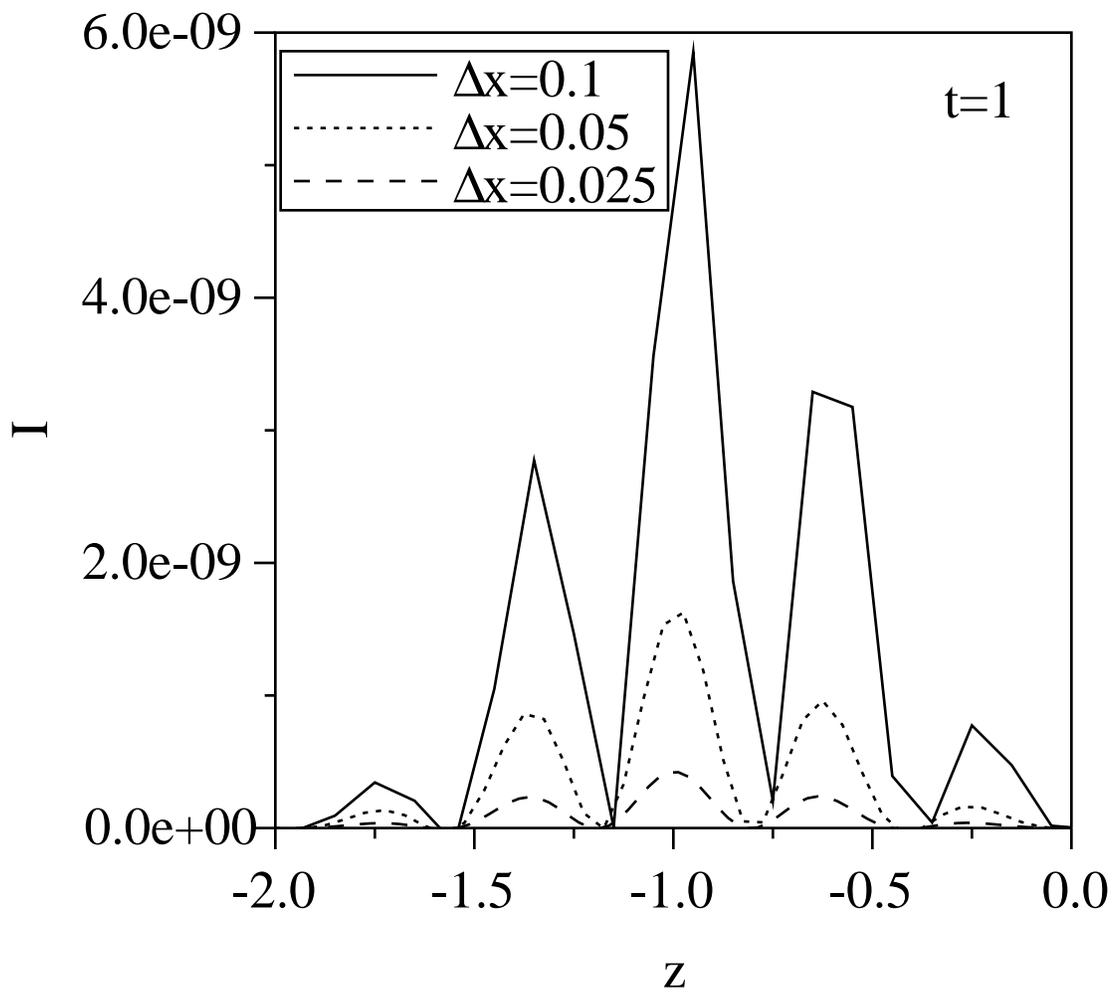
\begin{figure}
\caption{The curvature invariant $I$ is plotted for plane wave
evolutions.  It is known that all curvature invariants are zero for
plane wave spacetimes, and in this figure we see $I$ converging to zero
as we increase the resolution.}
\label{fig:curv_pw}
\end{figure}

\begin{figure}
\caption{The metric function $g_{xx}$ is shown for two plane waves
with the same parameters as the single wave packet, except a larger
amplitude $A_{L}=A_{R}=0.025$, and centered at $z=\pm3$.  In (a) we show the
evolution with the G code, and a fully nonlinear evolution.  Note the
drifting that takes place in the region where the waves collide.  In
(b) we show the same initial data now evolved with the linear
evolution equations.  No drifting is present when the linear evolution
equations are used.  In (c) we show the same initial data evolved with
the full nonlinear evolution but now with a MacCormack-like finite
differencing scheme.  The drifting is now greatly reduced.  In (d) we
evolve the initial data with the H code, and the drifting is similar
to that found in (c).  In (e) we again use the nonlinear G code, but
now with higher resolution compared to (a).  We find  that the drifting
decreases with resolution.  In general we find the drifting is a
nonlinear effect, that depends on the resolution and
finite-differencing scheme used.  }
\label{fig:colliding} 
\end{figure}

\begin{figure}
\caption{A scalar field $f$, evolved with a nonlinear wave equation
with nonlinear terms similar to those found in the Einstein equations
is shown.  The shape of the wave packets is similar to those used in
the collision of two waves in the previous section.  A staggered
leapfrog scheme is used.  In (b) we show a blow up of the region of
interaction to show the drifting. }
\label{fig:model} 
\end{figure}

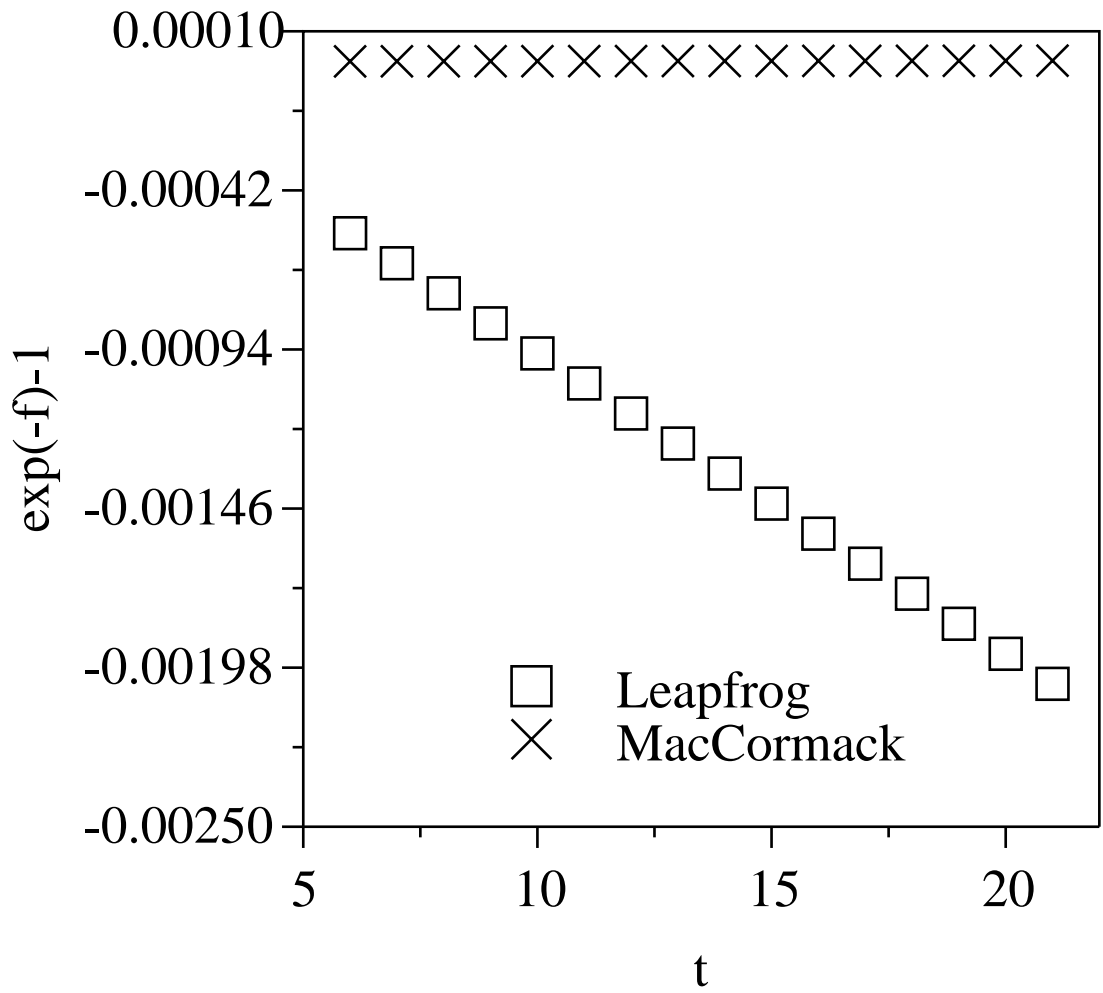
\begin{figure}
\caption{The quantity $e^{-f} - 1$ is shown plotted against time for
both the leapfrog and MacCormack-like schemes.  This shows
a solution of the form $f(t) = - \ln(c_1 t+ c_2)$ 
being excited by numerical error.  The constants $c_1$ and $c_2$ are
measured from the graph and depend on the resolution and numerical
scheme used.  The MacCormack scheme has a much smaller drift rate with
$c_1=1.4$x$10^{-7}$ and $c_2=1.0$, as compared to the leapfrog scheme
with $c_1=-9.8$x$10^{-5}$ and $c_2=1.0$.  These results were obtained
with $\Delta x=0.05$.  We see that with both these methods 
this solution converges away with increased resolution.}
\label{fig:modeldrift} 
\end{figure}

\begin{figure}
\caption{The metric function $g_{xx}$ is shown for a small 
amplitude Teukolsky wave with azimuthal mode number $m=0$.  
This corresponds to a perturbation in the metric function of about 
$0.025\%$.  In (a) we show the evolution using the G code.  In (b) we 
magnify the region near the axis showing the wave in the metric 
function which rapidly drops off as the wave travels outward.  In (c) 
we show the evolution with the H code.}
\label{fig:quadrupole_small} 
\end{figure}

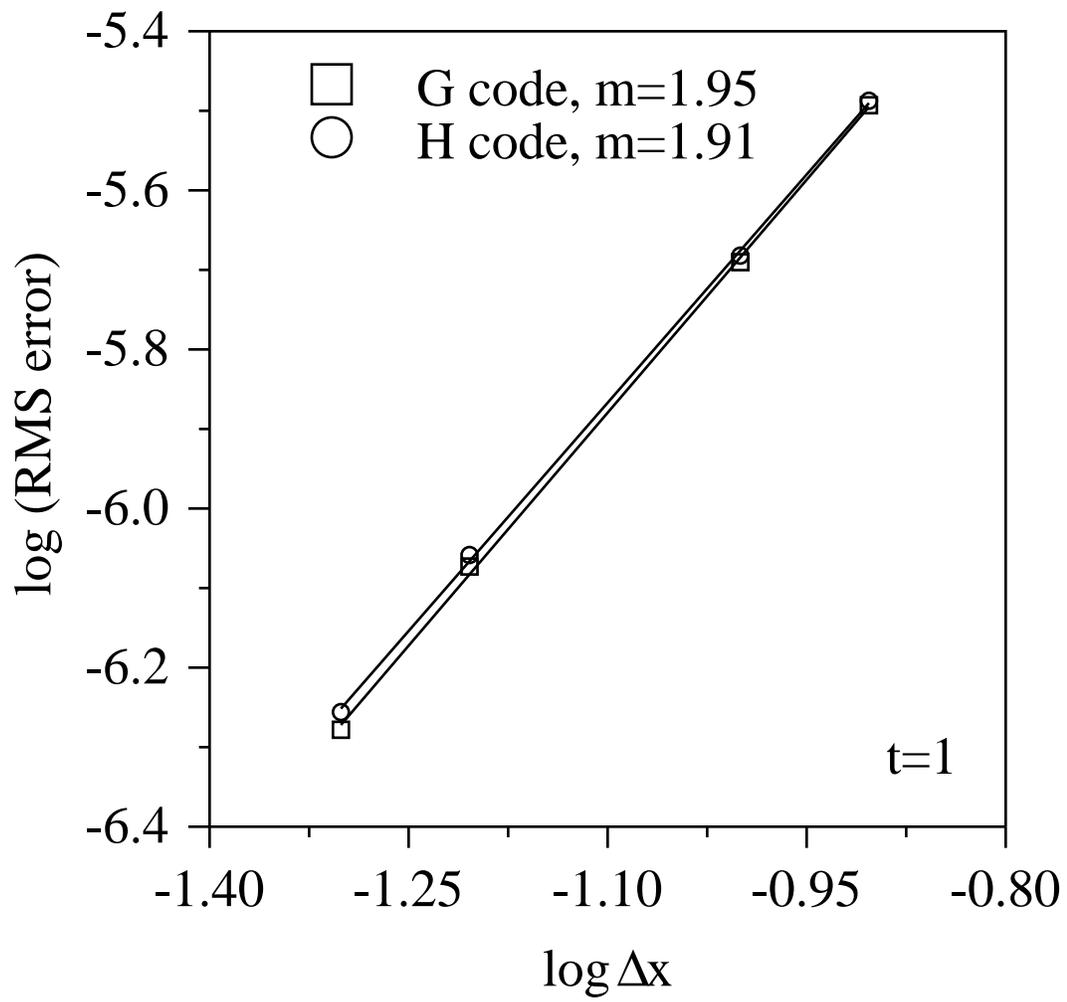
\begin{figure}
\caption{The log of the RMS error is again plotted as in
Fig.~3, to test the convergence of the code.  At $t=1$ we get
a convergence rate of 1.95 for the G code, and 1.91 for the H code.}
\label{fig:errortw} 
\end{figure}

\begin{figure}
\caption{The metric function $g_{xx}$ is shown for a moderate
amplitude Teukolsky wave with azimuthal mode number $m=0$.
This corresponds to an initial perturbation in the metric function of about
$0.5\%$.  The early part of the evolution is virtually identical to
Fig.~9a, but at late times, after the wave has dispersed out, we now
see a dipping in the metric function near the origin.  In (a) the
initial data is evolved with the full nonlinear 
equations using the G code, and we clearly see the late time dipping.
In (b) we evolve with the linear evolution  
equations and see no evidence of the dipping.  In (c) we again use the
nonlinear evolution, but increase the resolution and find the dipping
does not converge away. }
\label{fig:quadrupole} 
\end{figure}

\begin{figure}
\caption{In (a) we show the conformal factor after using the Teukolsky
initial data and solving the IVP.  In (b) we show the evolution of
this initial data which now solves the constraint equation up to
numerical error.  We see that the evolution of this data is virtually
the same as in Fig.11a and the dipping of the metric
function is still present.  }
\label{fig:conf} 
\end{figure}

\begin{figure}
\caption{The effect of slicing on the 
evolution of the quadrapole wave + IVP initial data.  In (a) we see the 
evolution of the metric function $g_{xx}$ with geodesic slicing.  
In (b) we show $\tau_{tt}$, the stress-energy pseudo-tensor.  In (c) 
we show the curvature invariant $I$. In (d) we show the same initial 
data now evolved with maximal slicing.  Note that there is no dipping 
in the metric function.  In (e) we again show $\tau_{tt}$, and in (f) 
the curvature invariant $I$. Note that even though the evolution of the 
metric function differs with slicing, $\tau_{tt}$ and $I$ remain the 
same which suggests that we are seeing a coordinate, rather than 
geometric, effect.  In (g) we show the lapse $\alpha$.  Note that 
$\alpha$ is very close to one.  It is the shape of the lapse rather than 
its size, that keeps the metric function from dipping near the origin. 
We also note that the pseudo-tensor and the invariants are not defined
in the first boundary cells in our computational domain, and so we arbitrarily
assign a value of zero to the left-most point in the graphs.}
\label{fig:slicing_tw} 
\end{figure}

\begin{figure}
\caption{We define a critical time $t_{critical}$ at which the RMS
relative difference of the linear evolution disagrees with nonlinear
evolution by $10\%$.  In (a) we look at $t_{critical}$ for the metric
function $g_{xx}$, and compare it against the size of initial
perturbation in the metric function $g_{zz}$.  Since the metrics show
the coordinate dipping, this is a measure of the onset of nonlinear coordinate
effects.  We find that there is an approximate power law dependence of
$t_{critical}$.  The error bars in this graph come from the fact that
the data is only analyzed in time intervals of $\Delta t=0.1$.  In (b)
we do the same, but for the curvature invariant $I$.  Since the
invariant is coordinate independent, it is a measure of nonlinear
geometric effects.  We find that nonlinear geometric effects occur at
a later time than nonlinear coordinate effects for the amplitudes
considered here.
}
\label{fig:t_crit}
\end{figure}

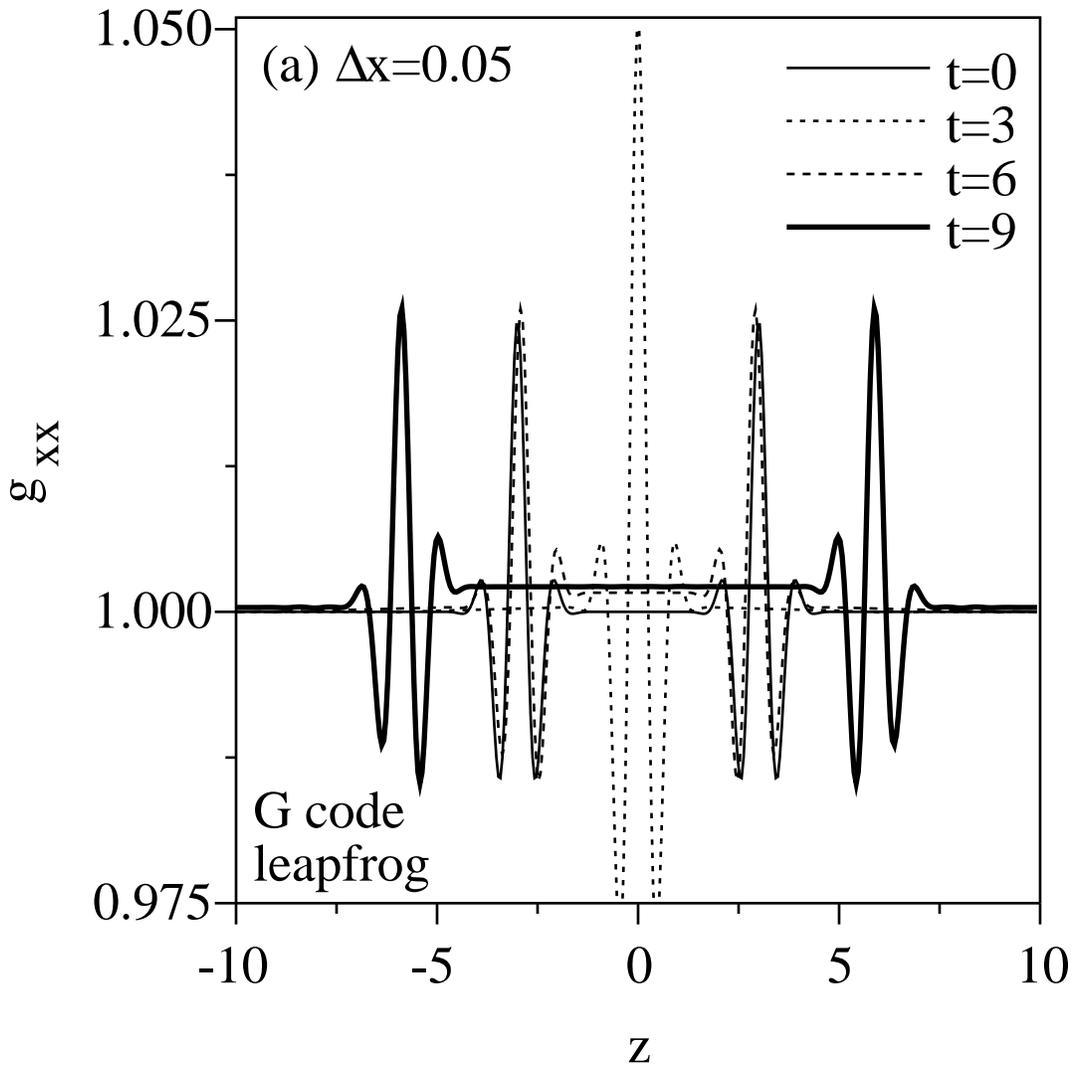
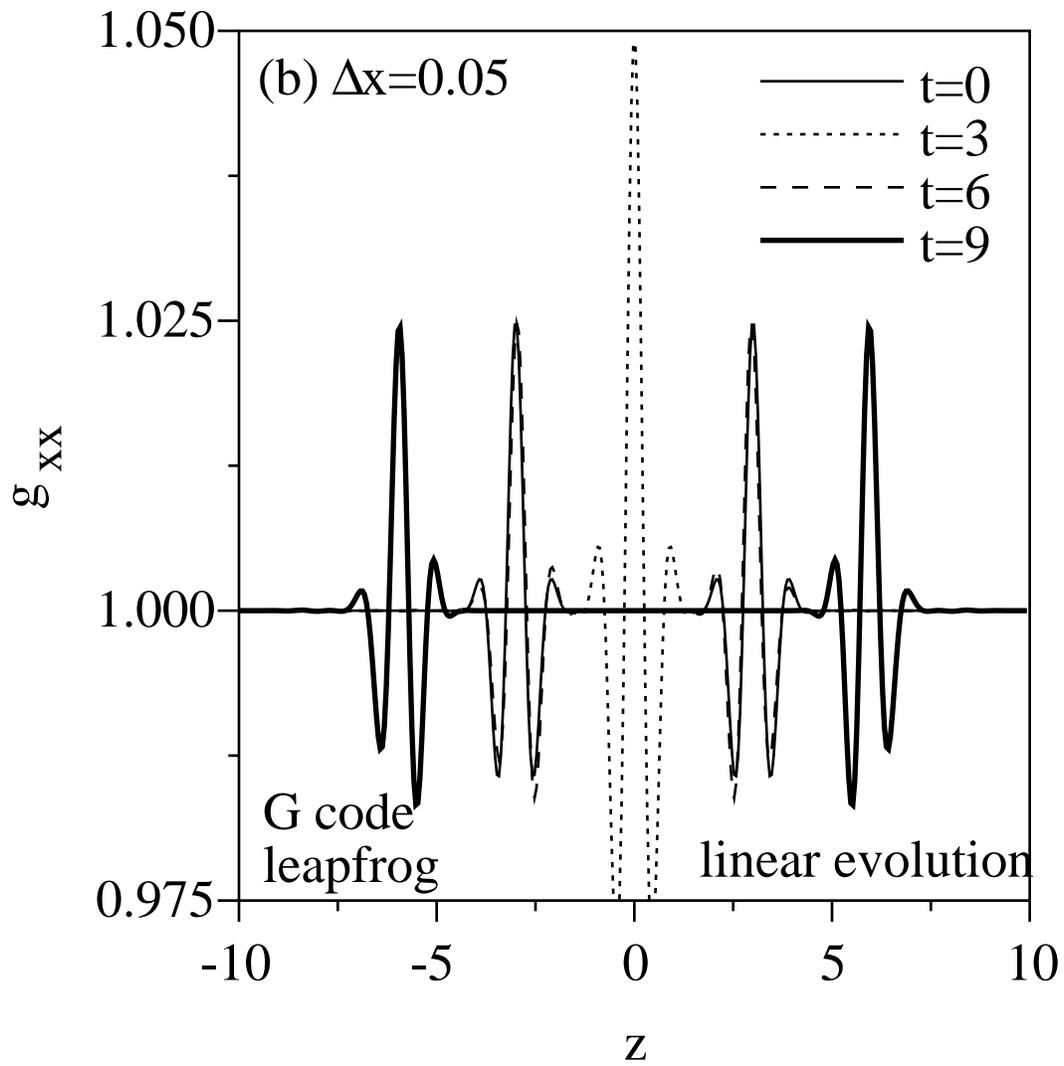
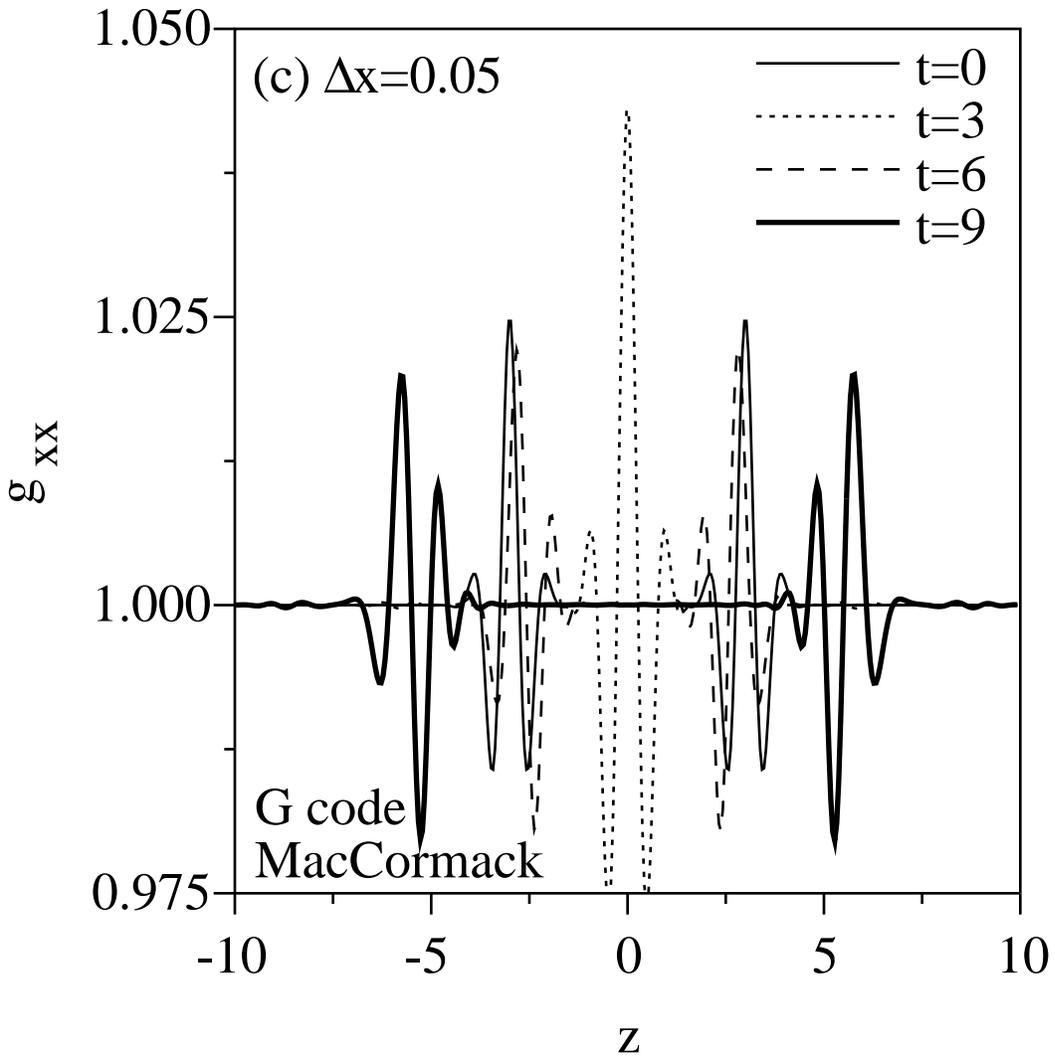
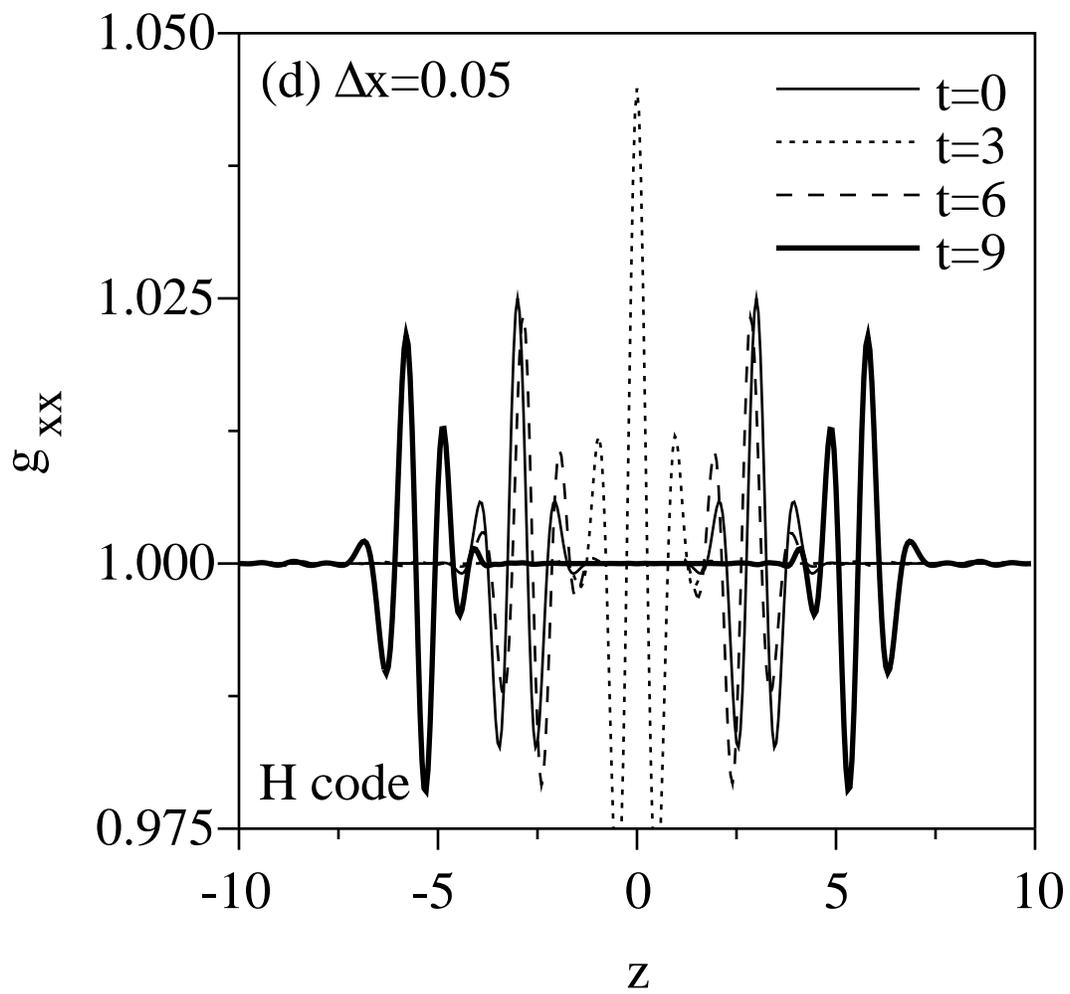
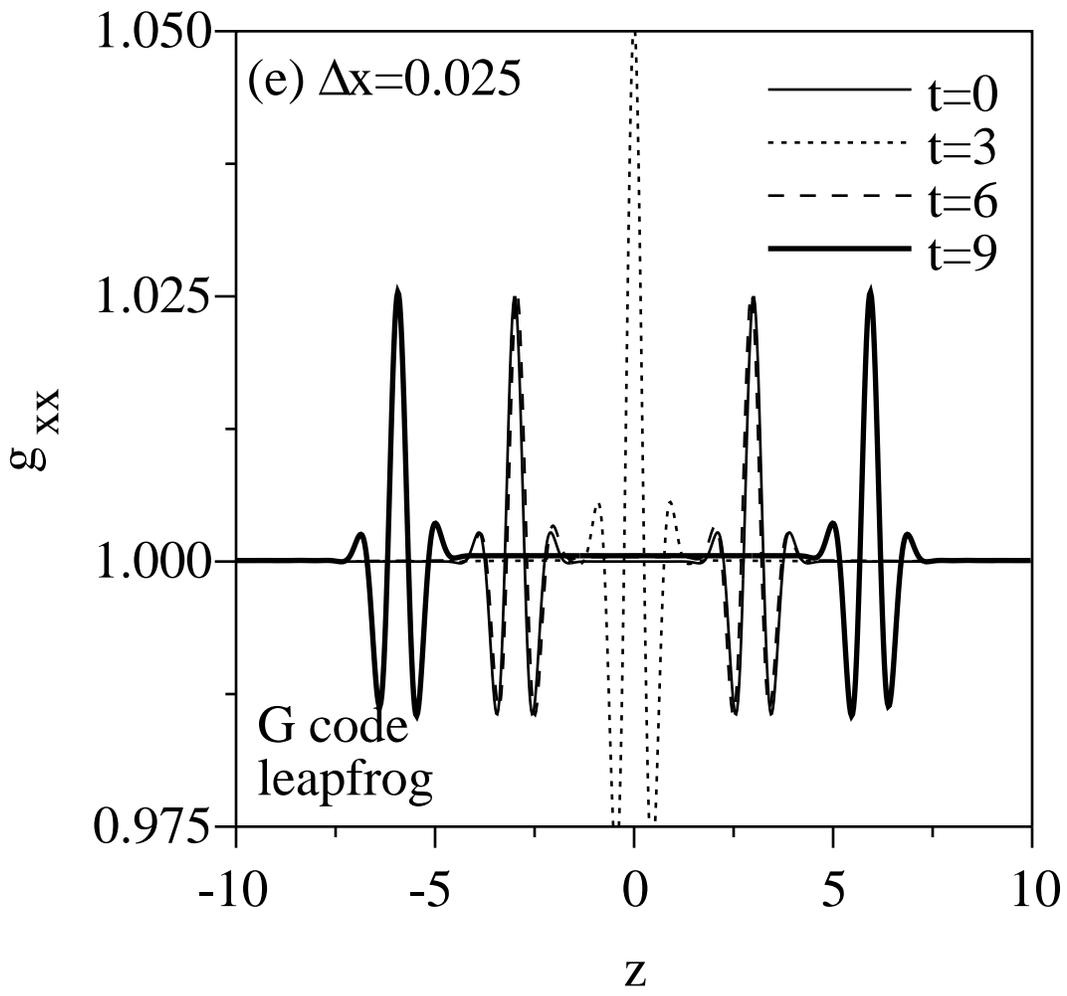
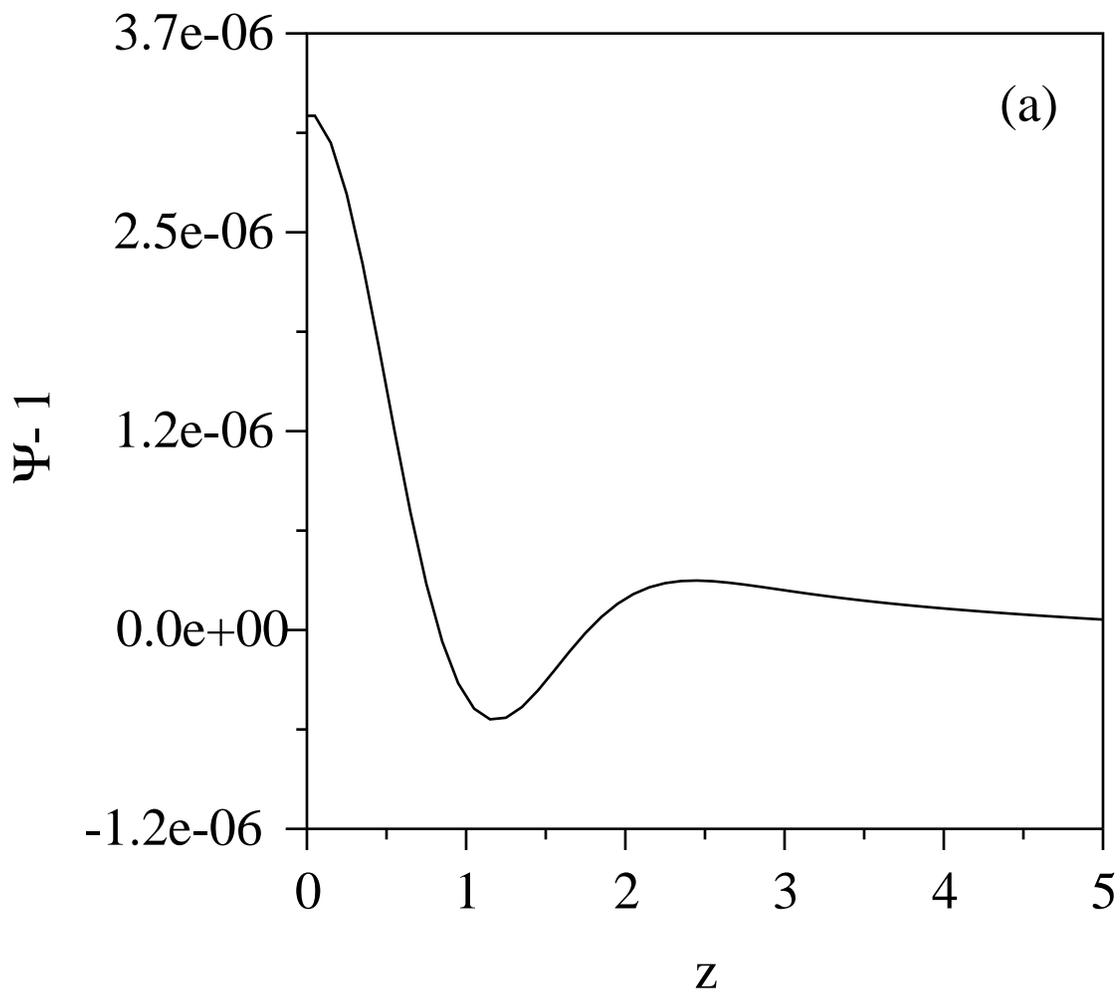
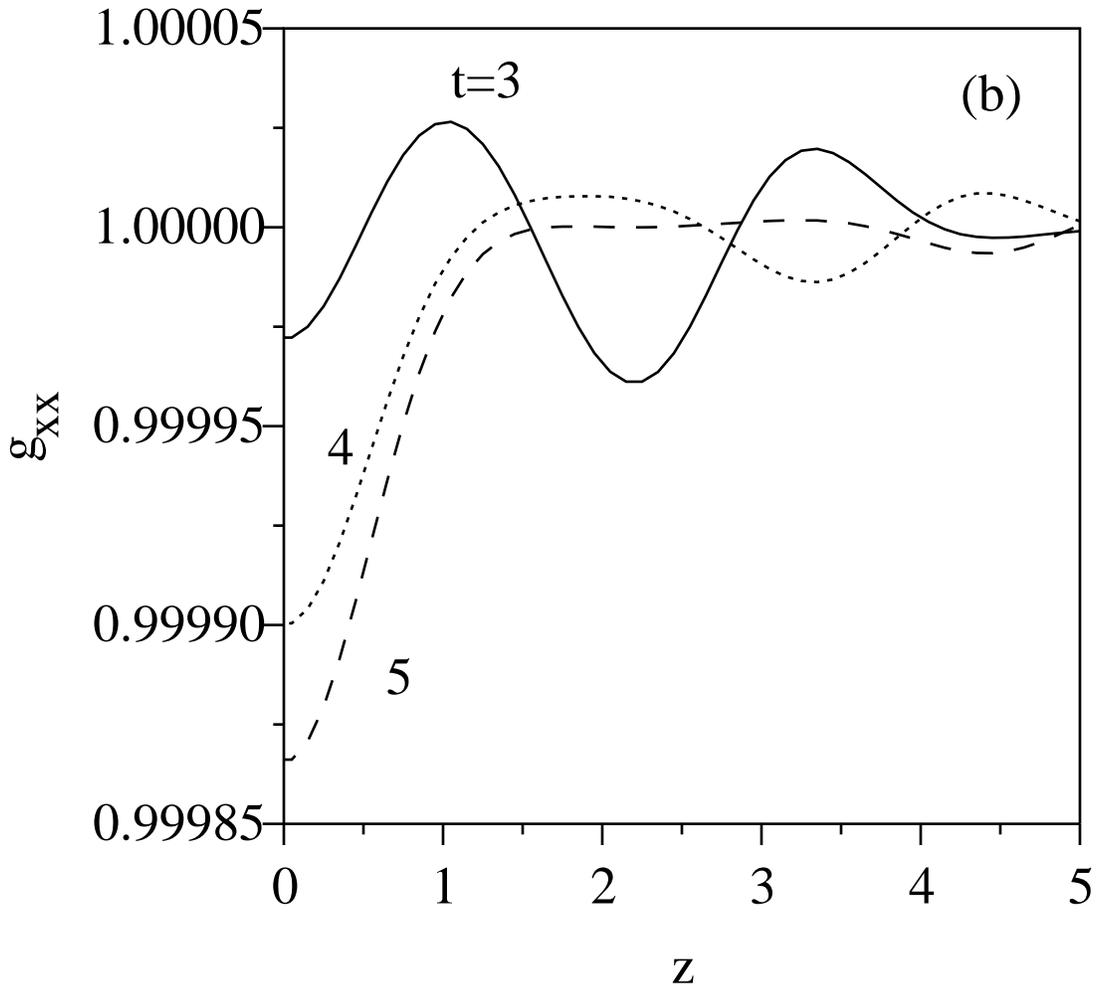
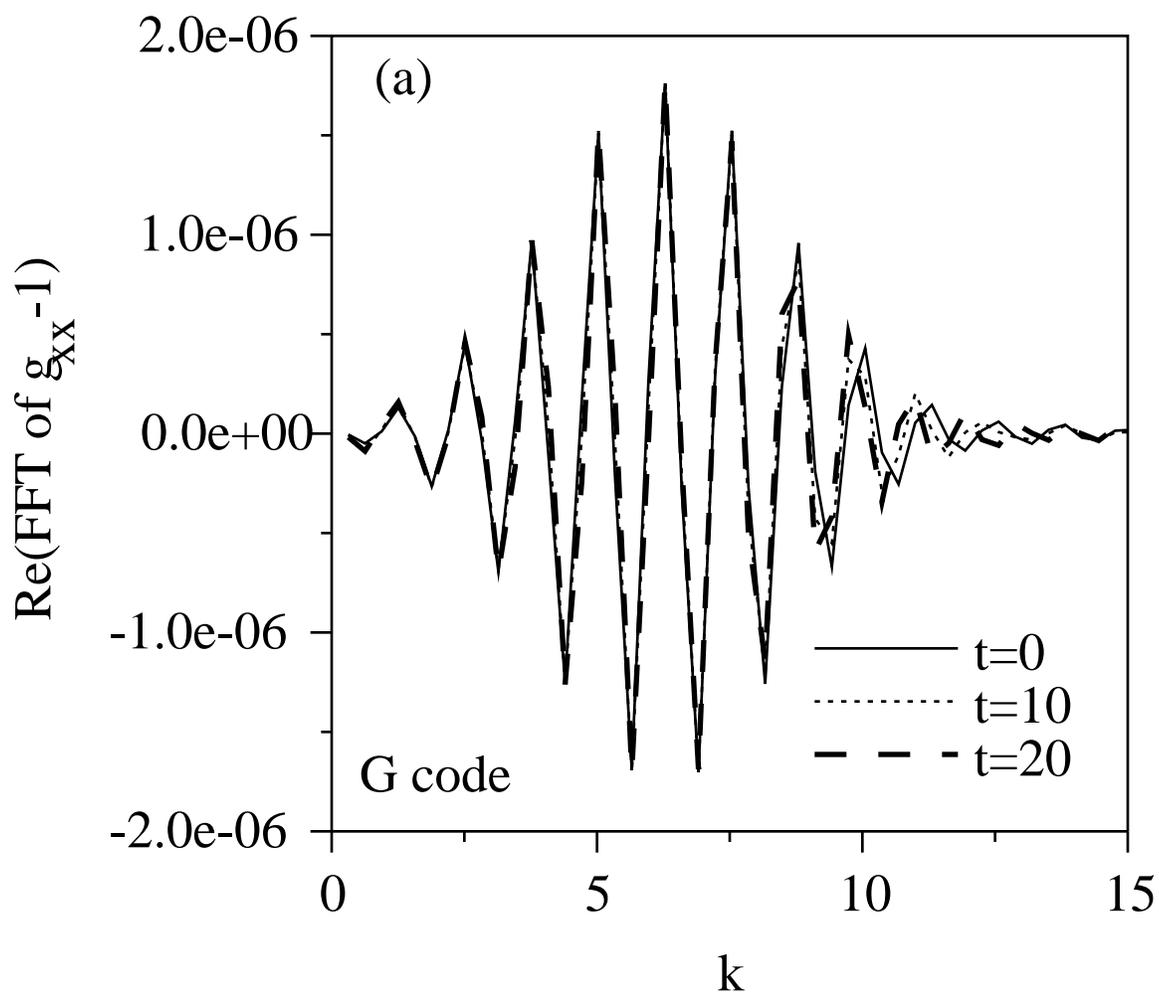
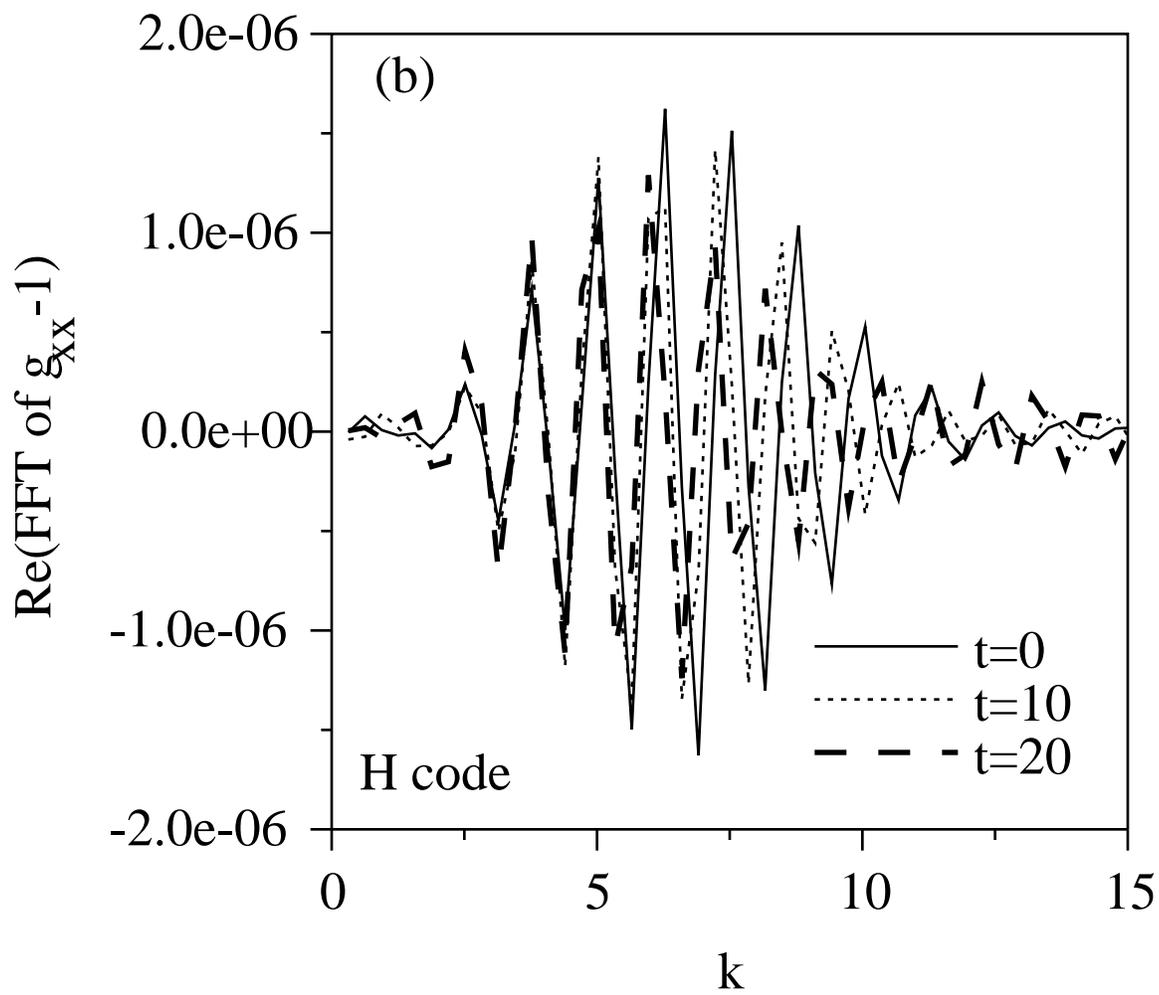
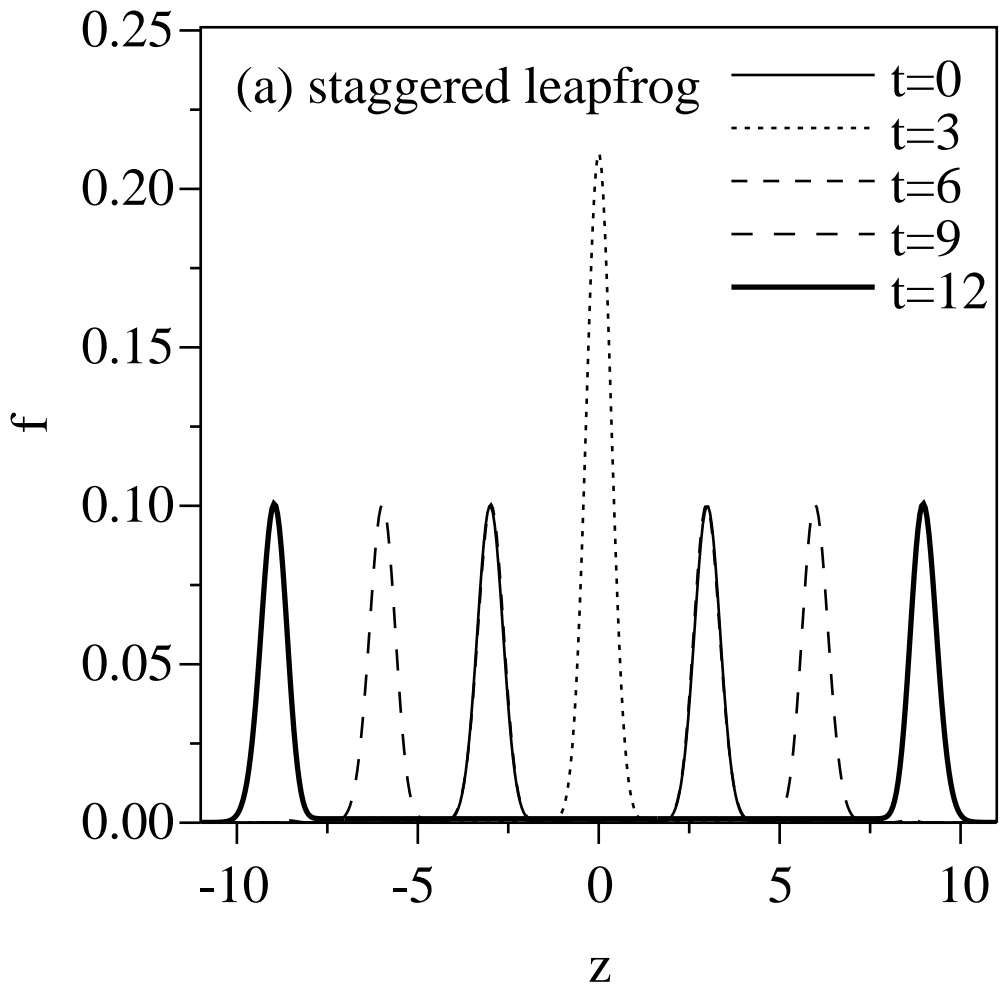
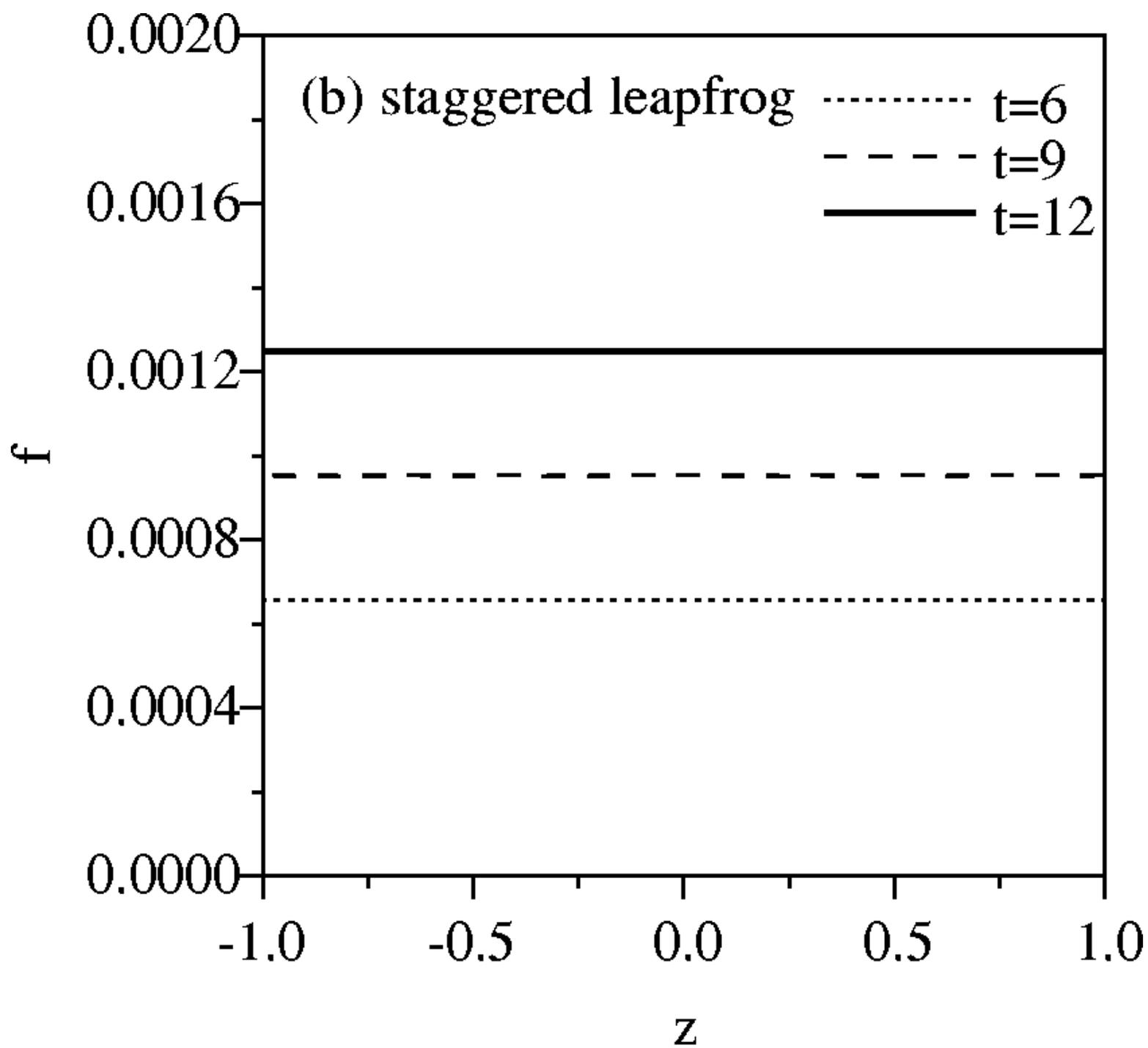
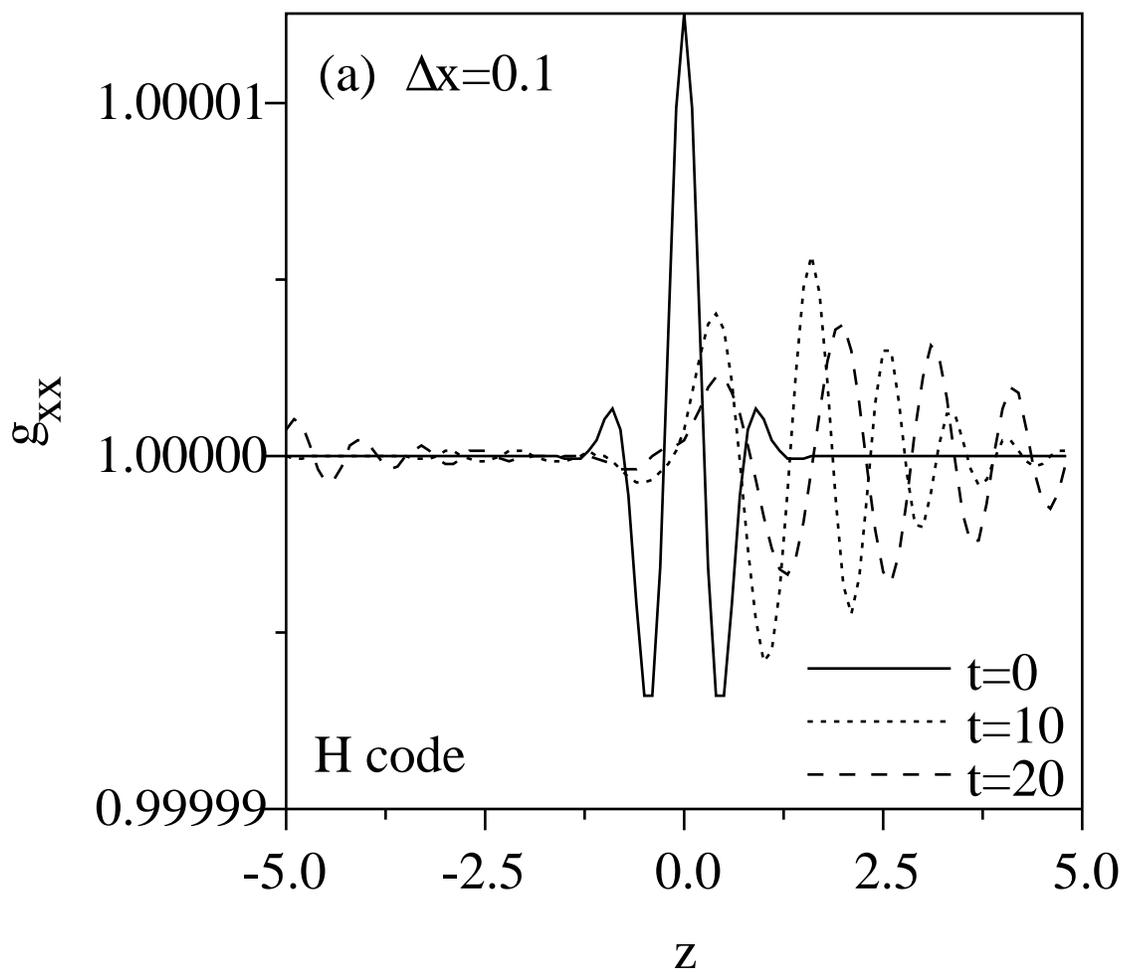
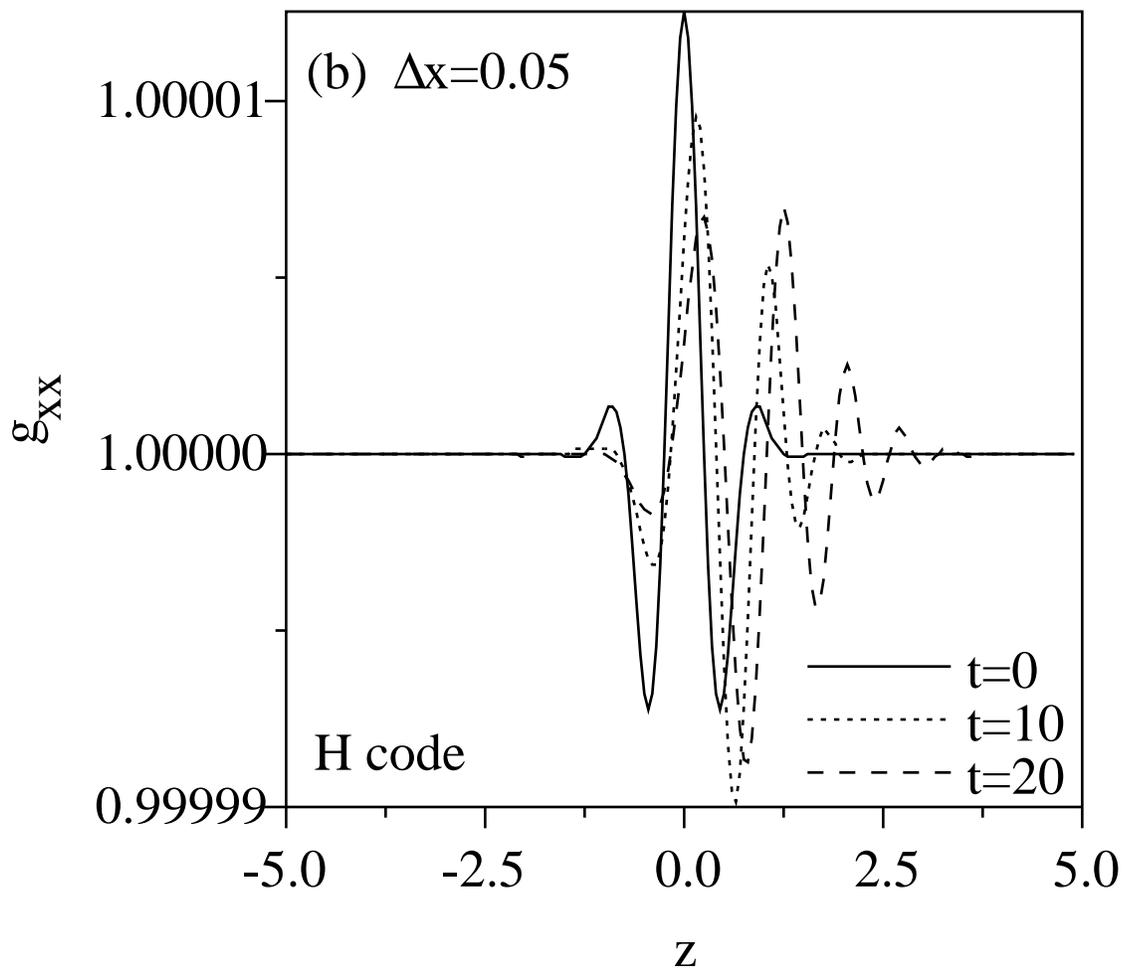
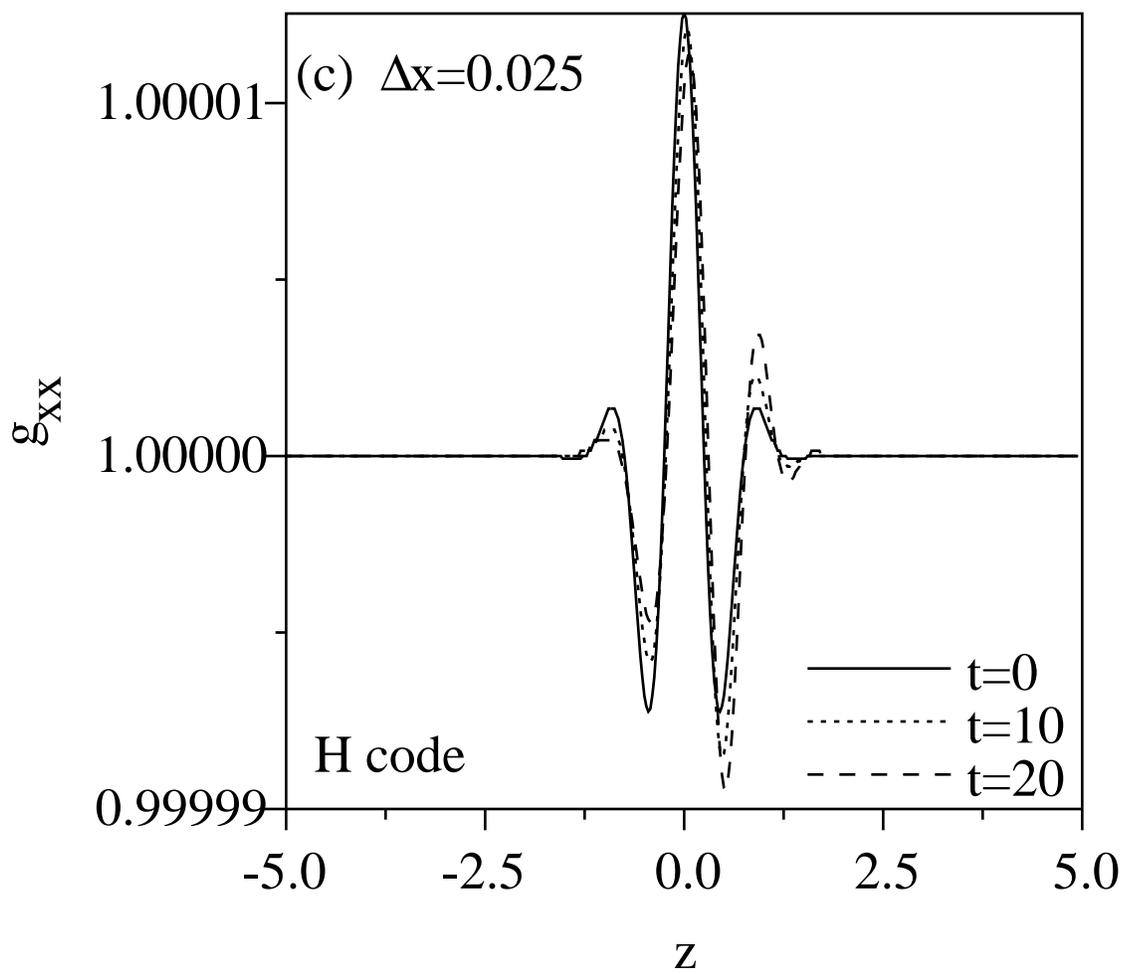
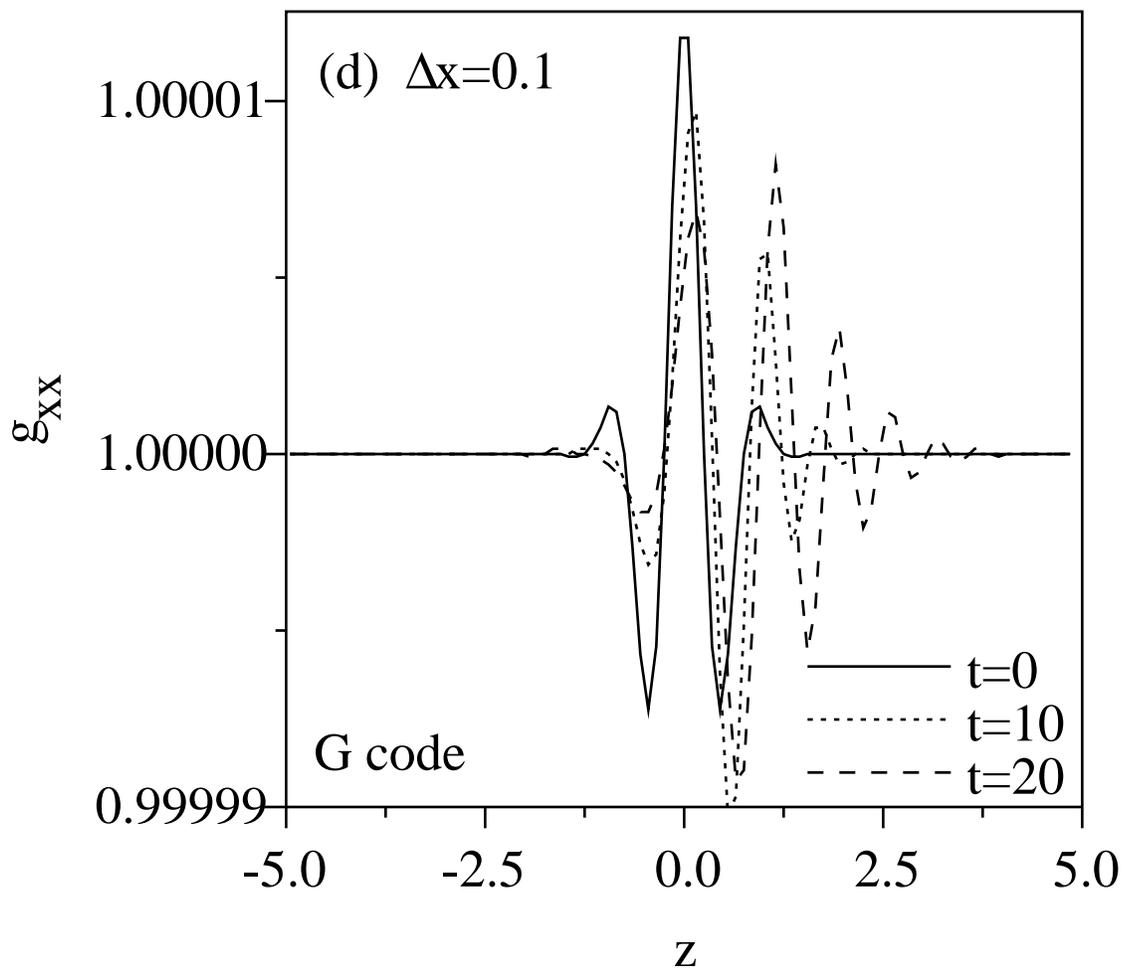
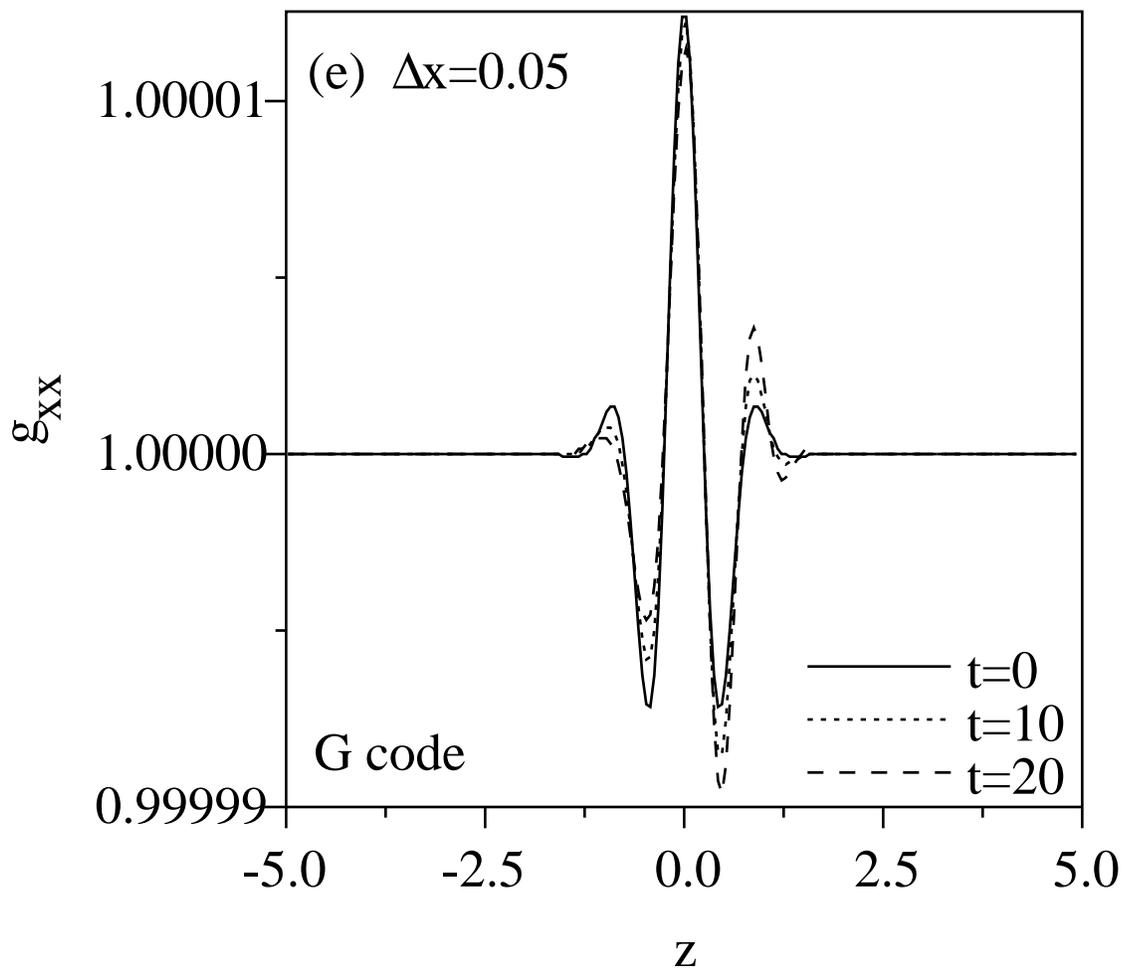
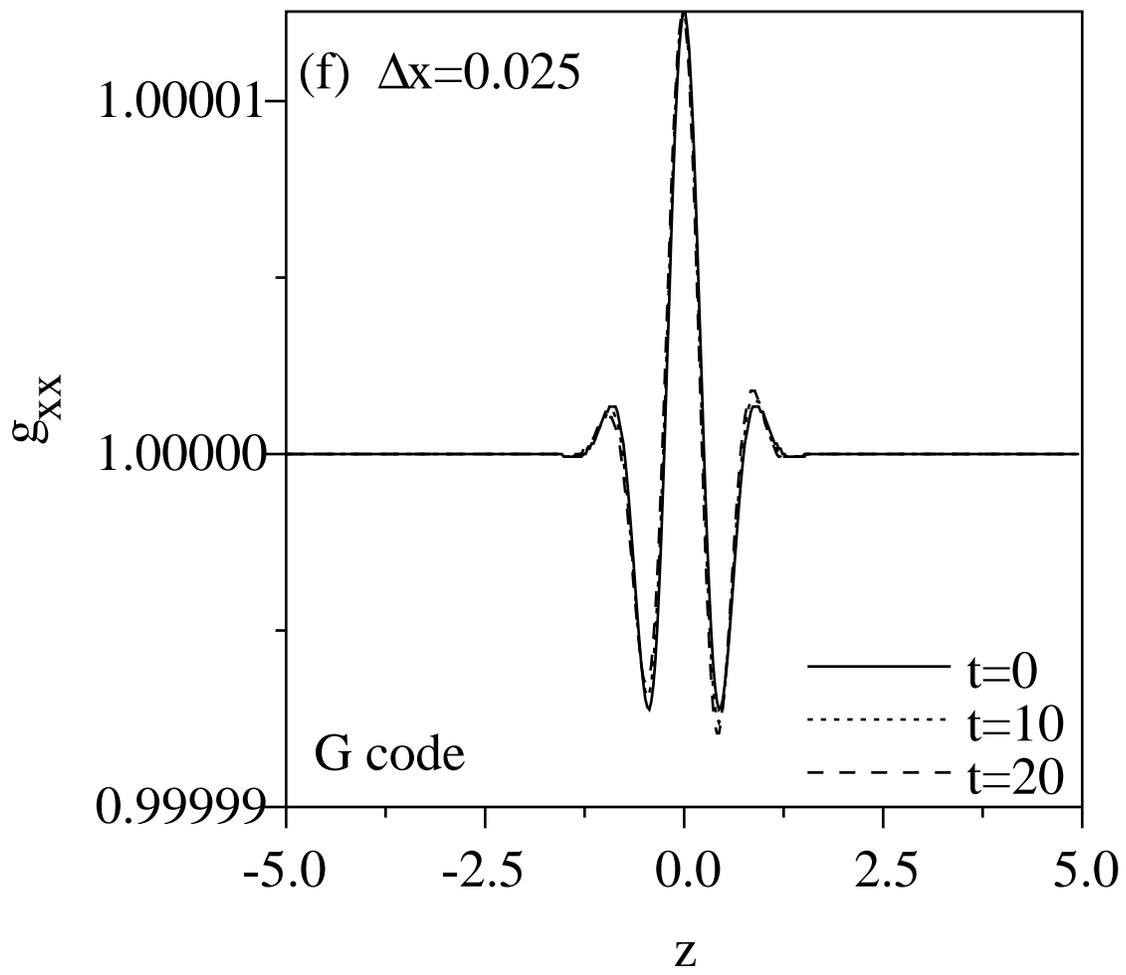
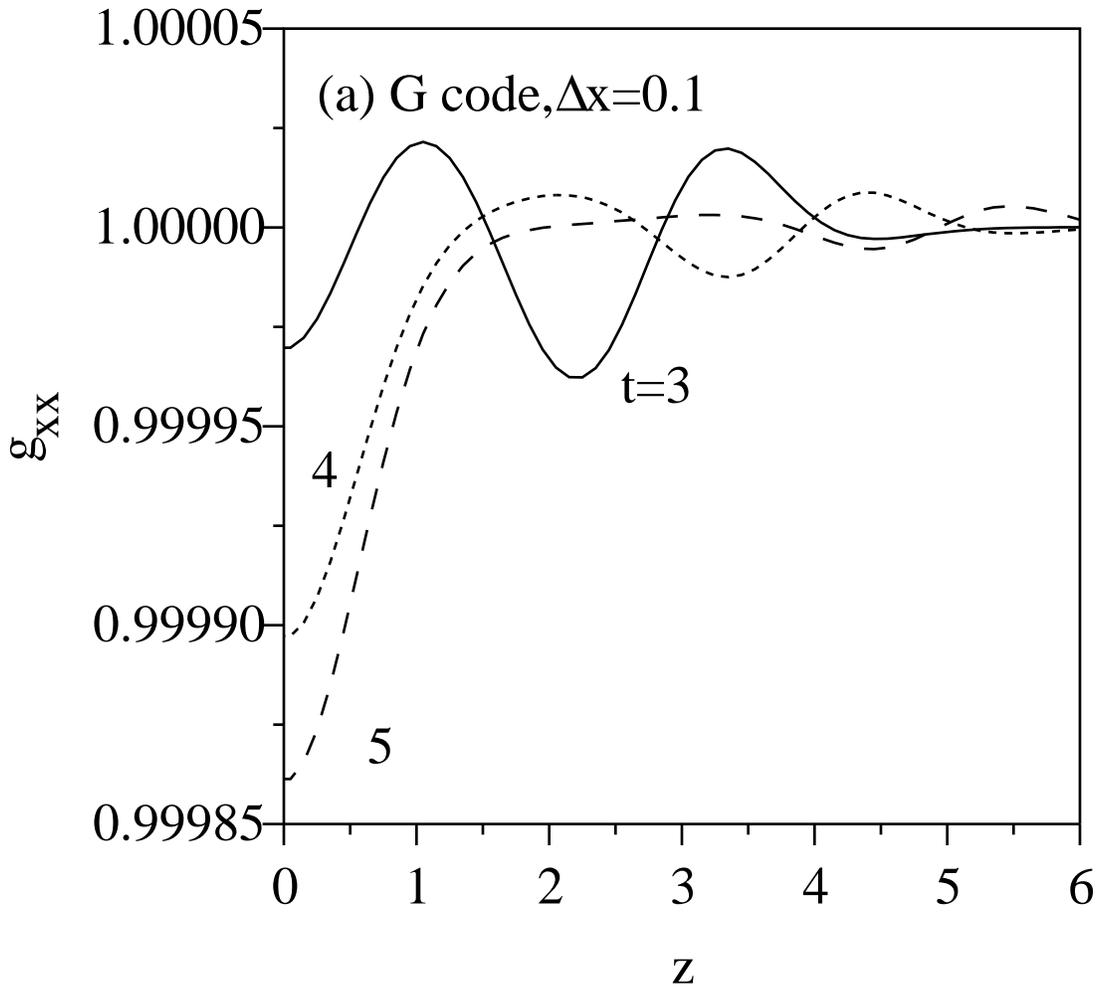
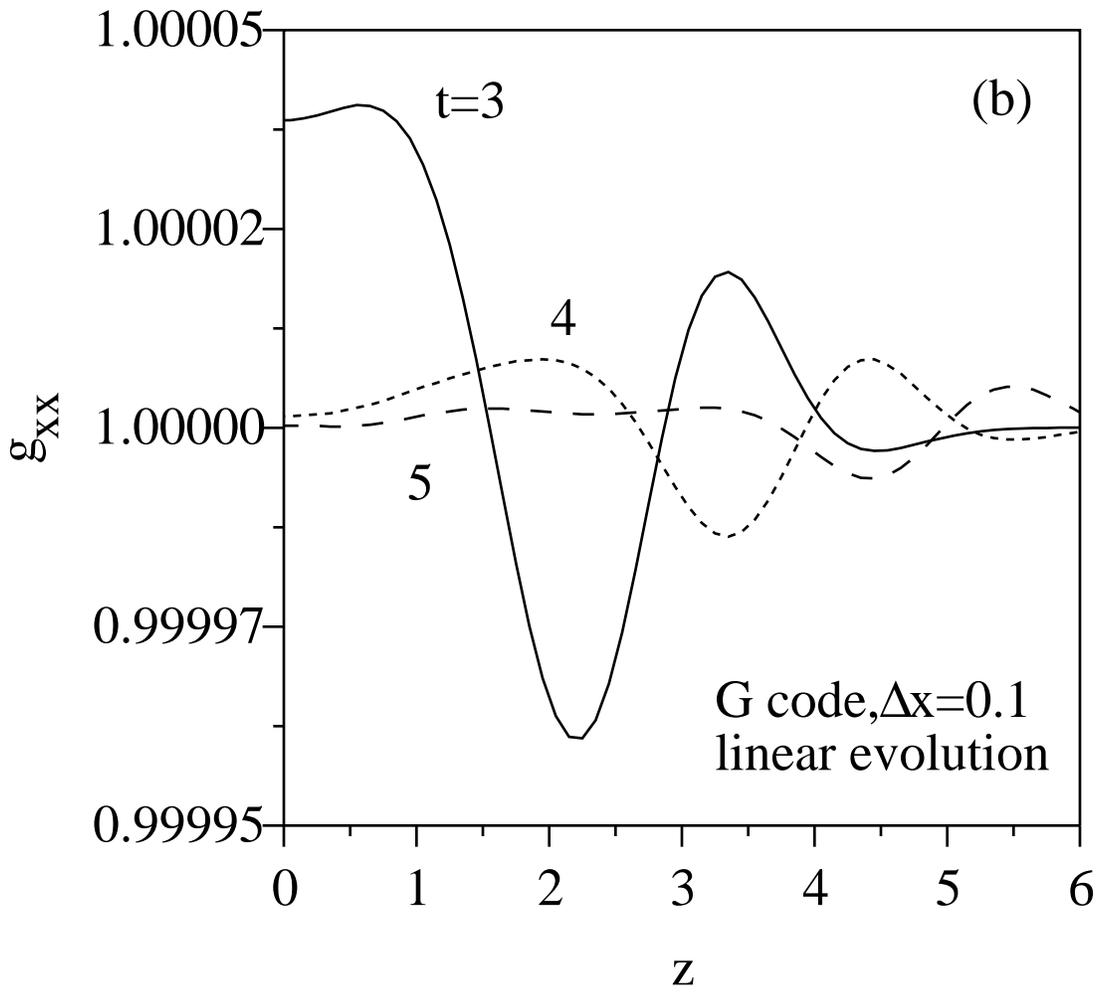
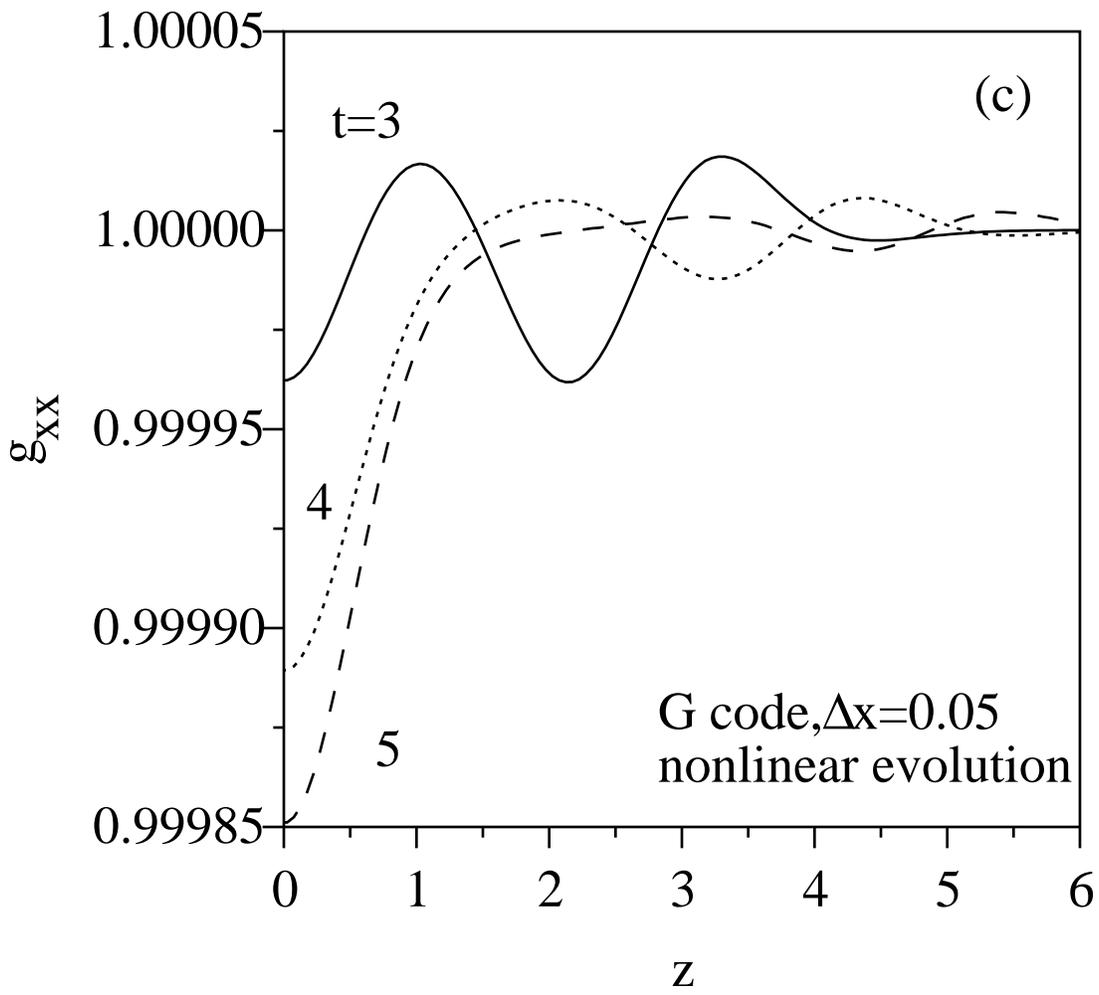
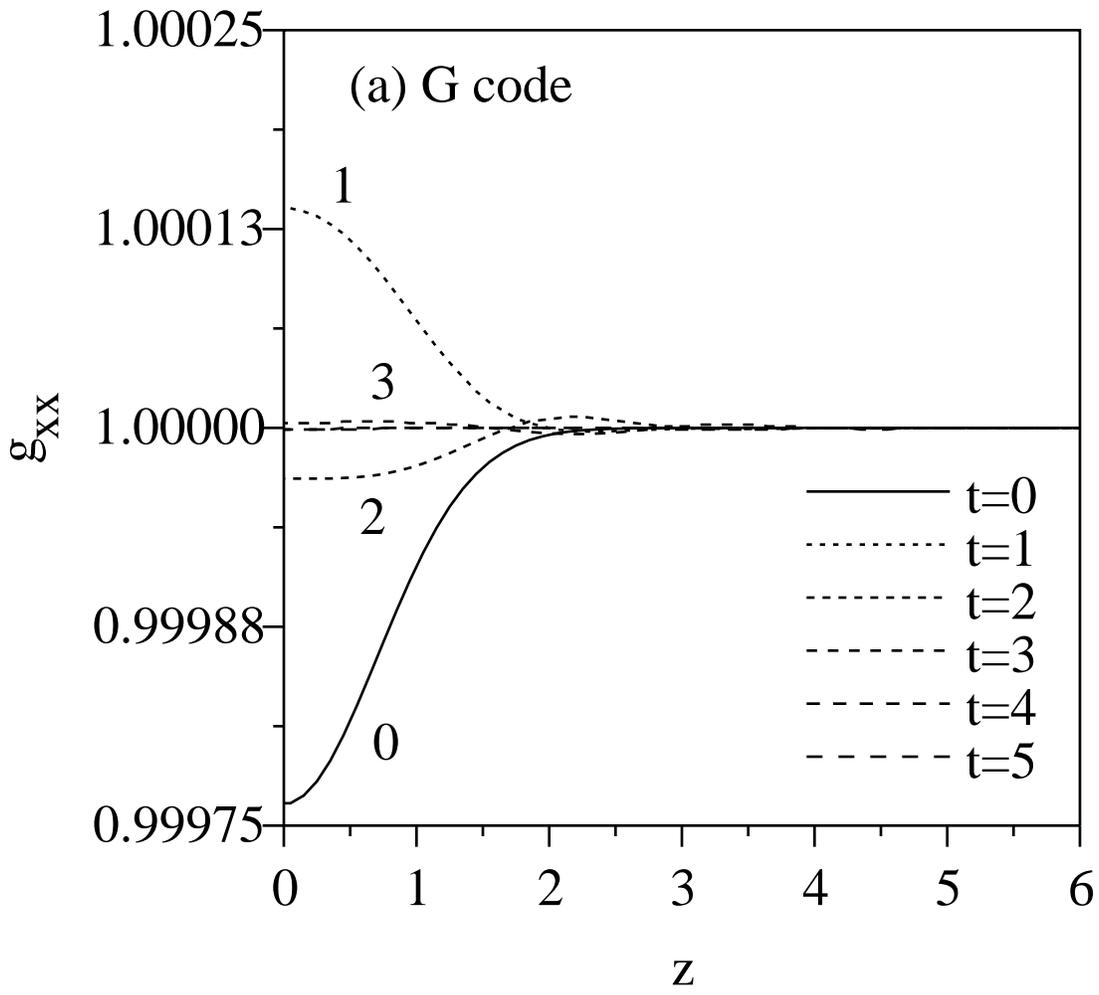
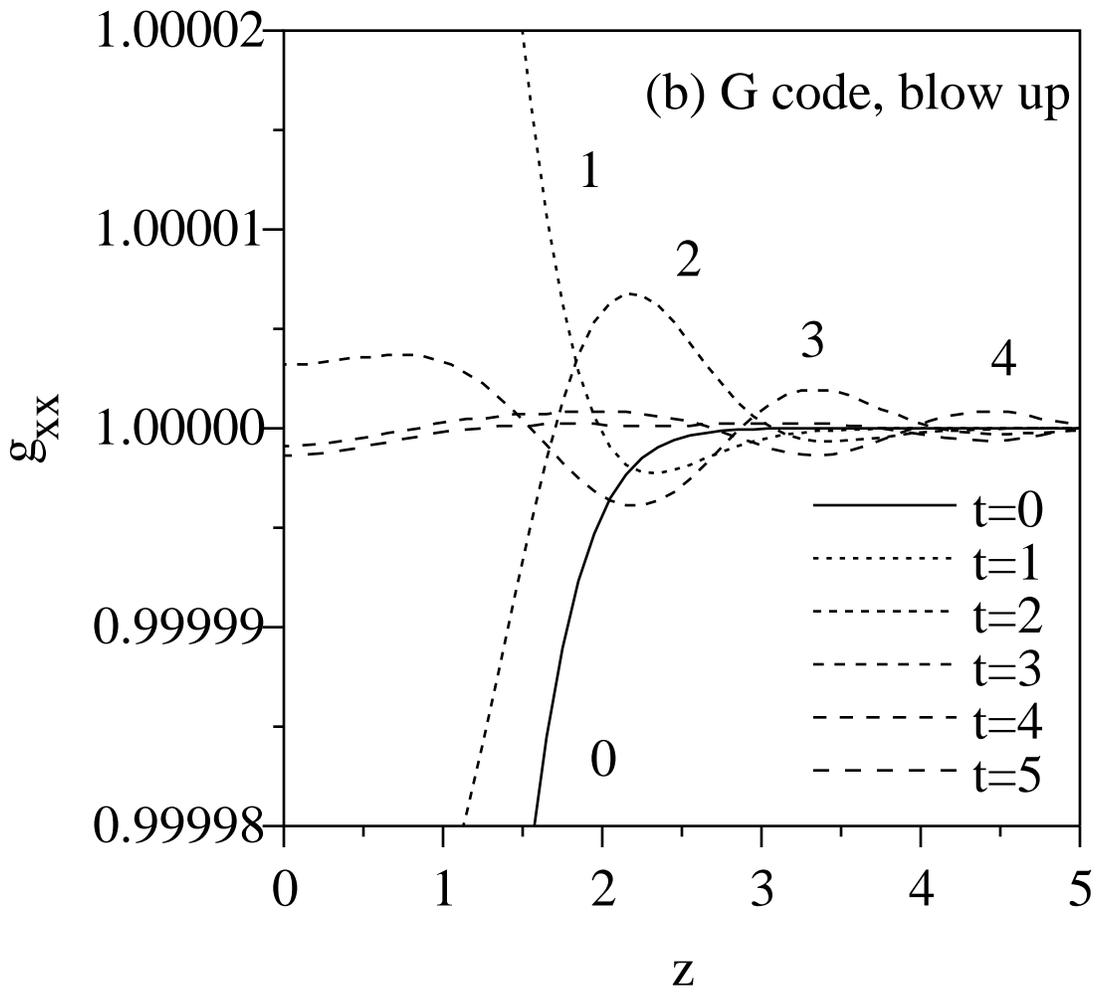
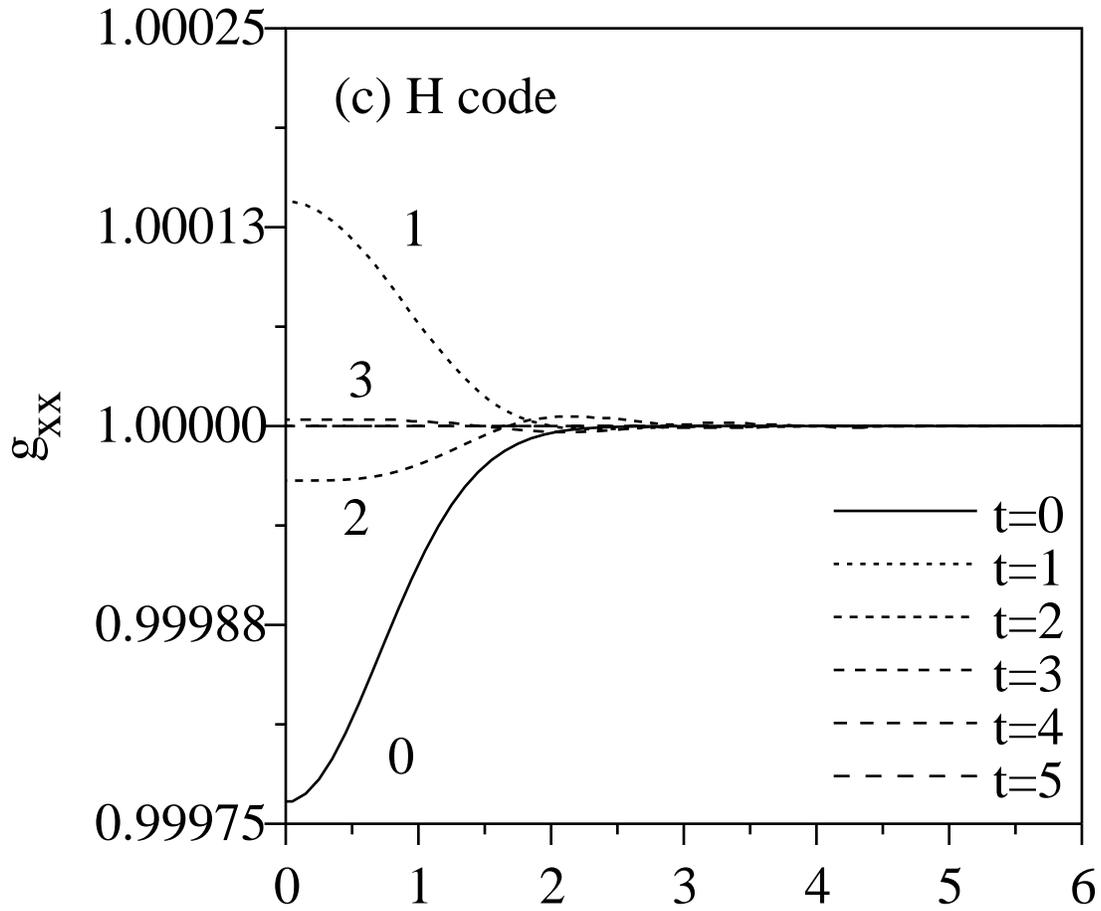
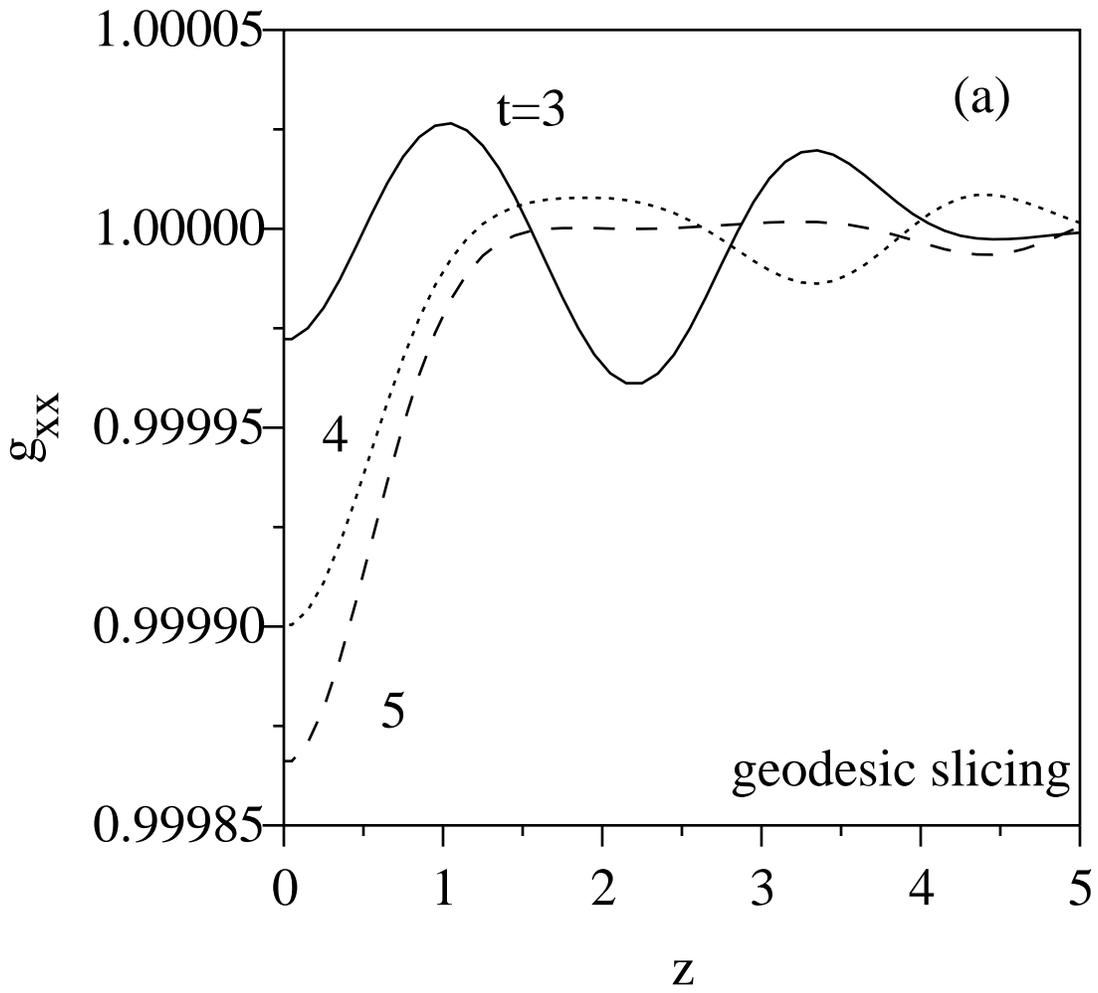
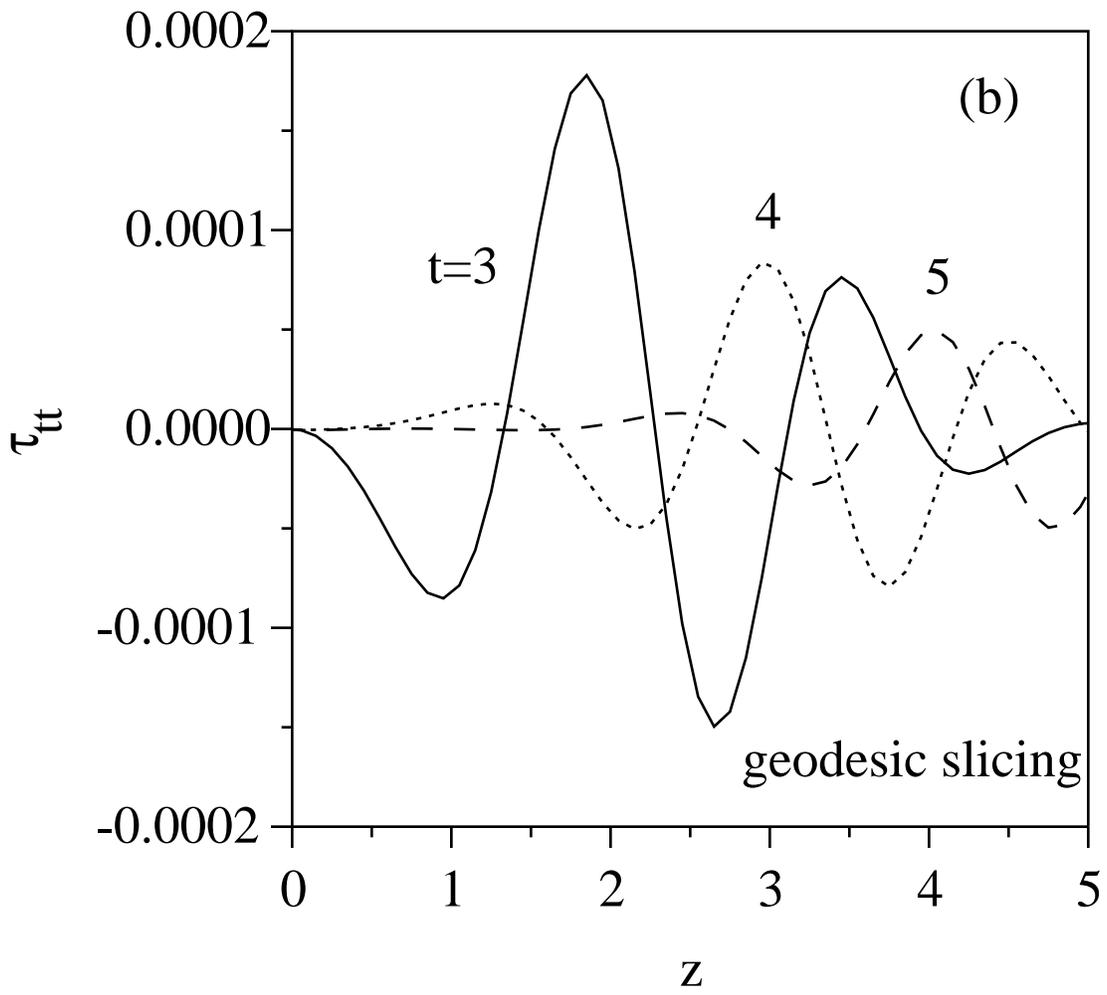
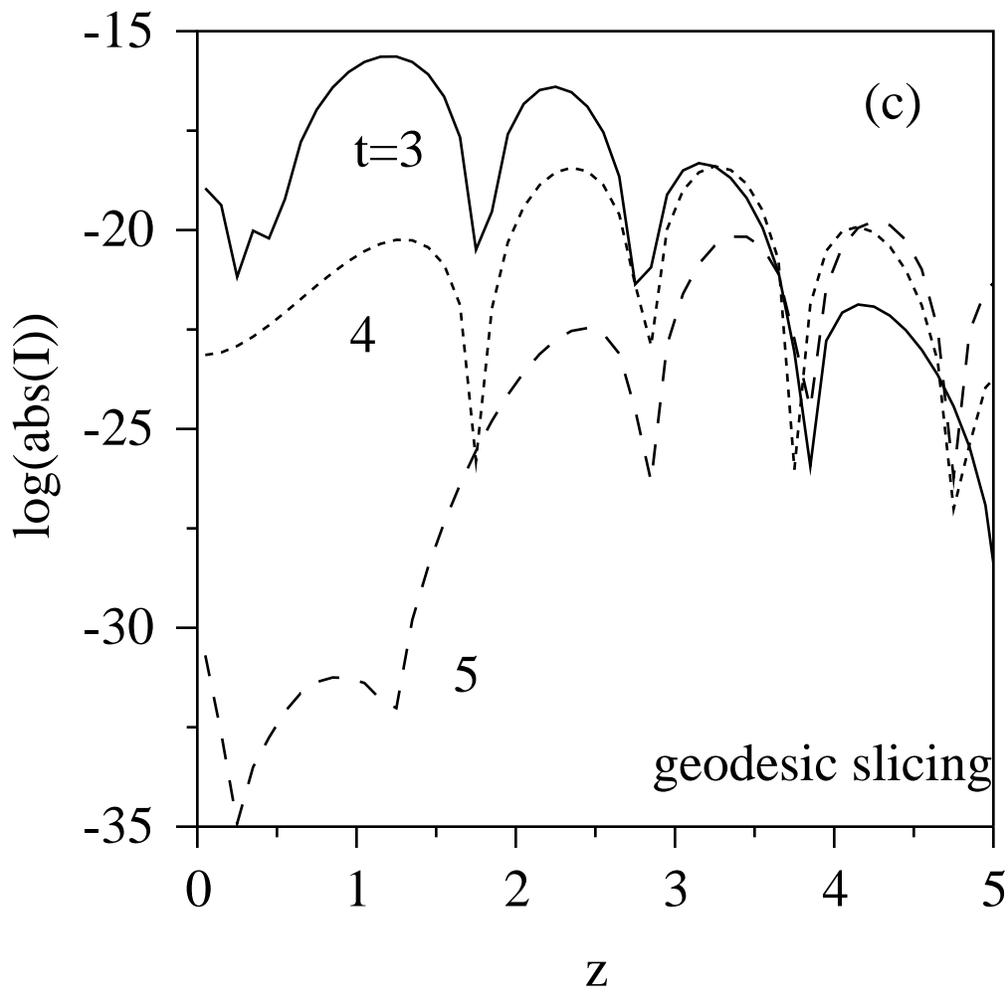
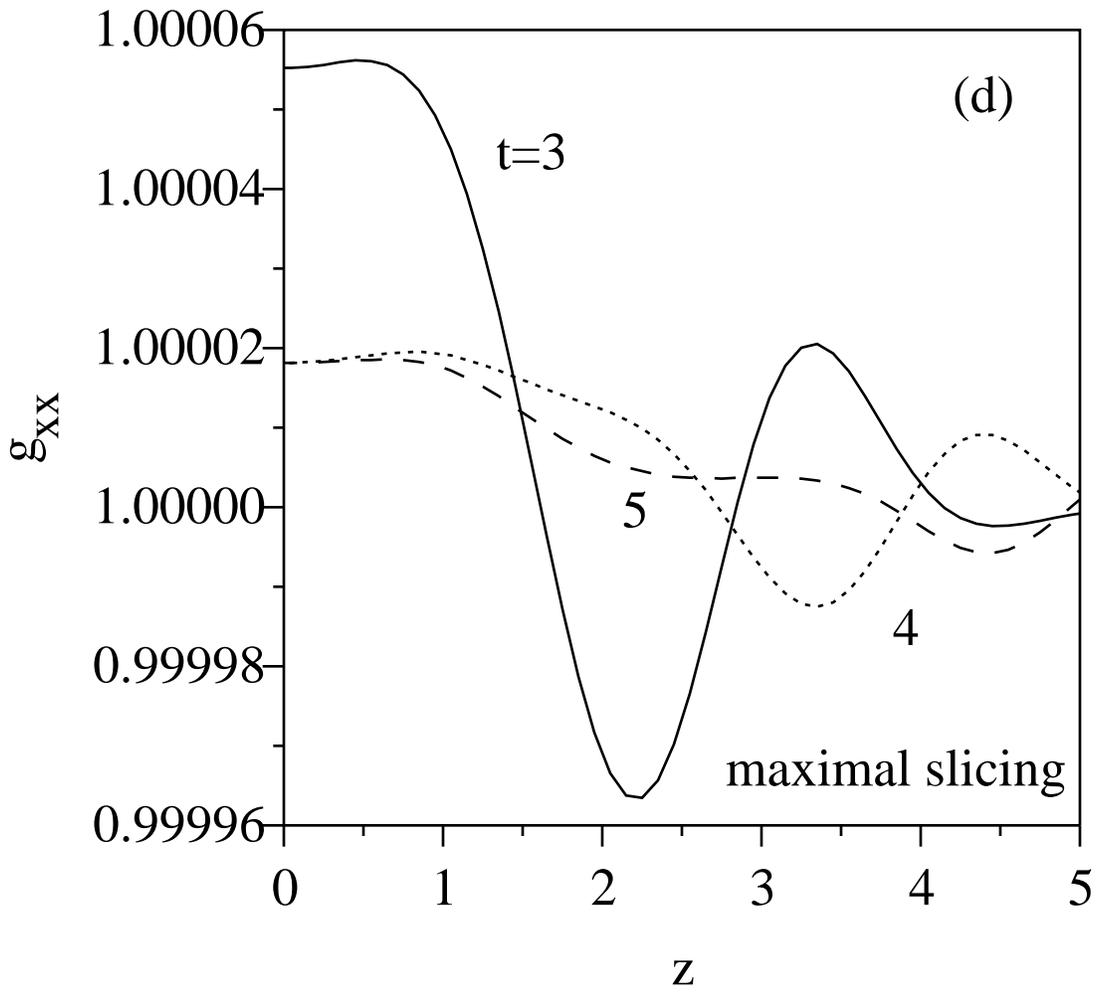
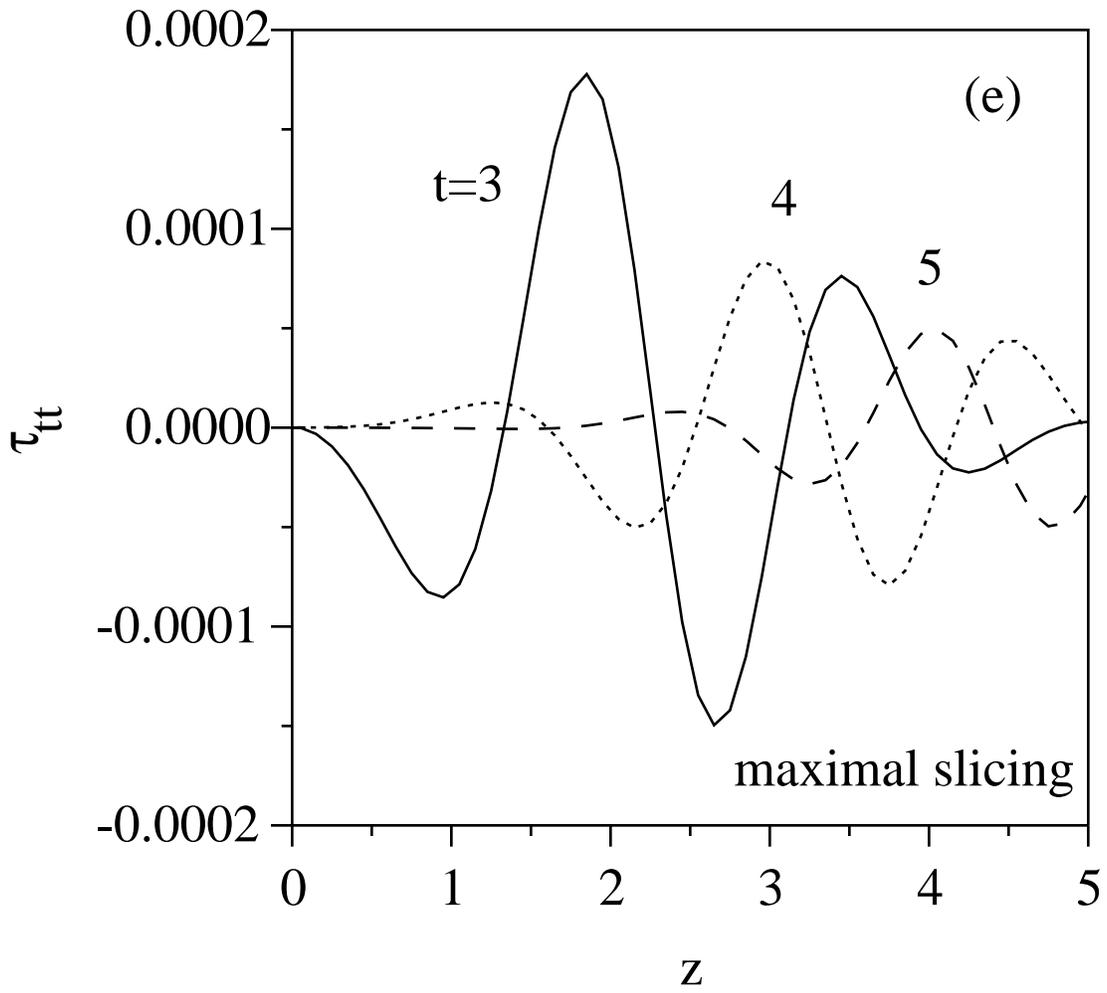
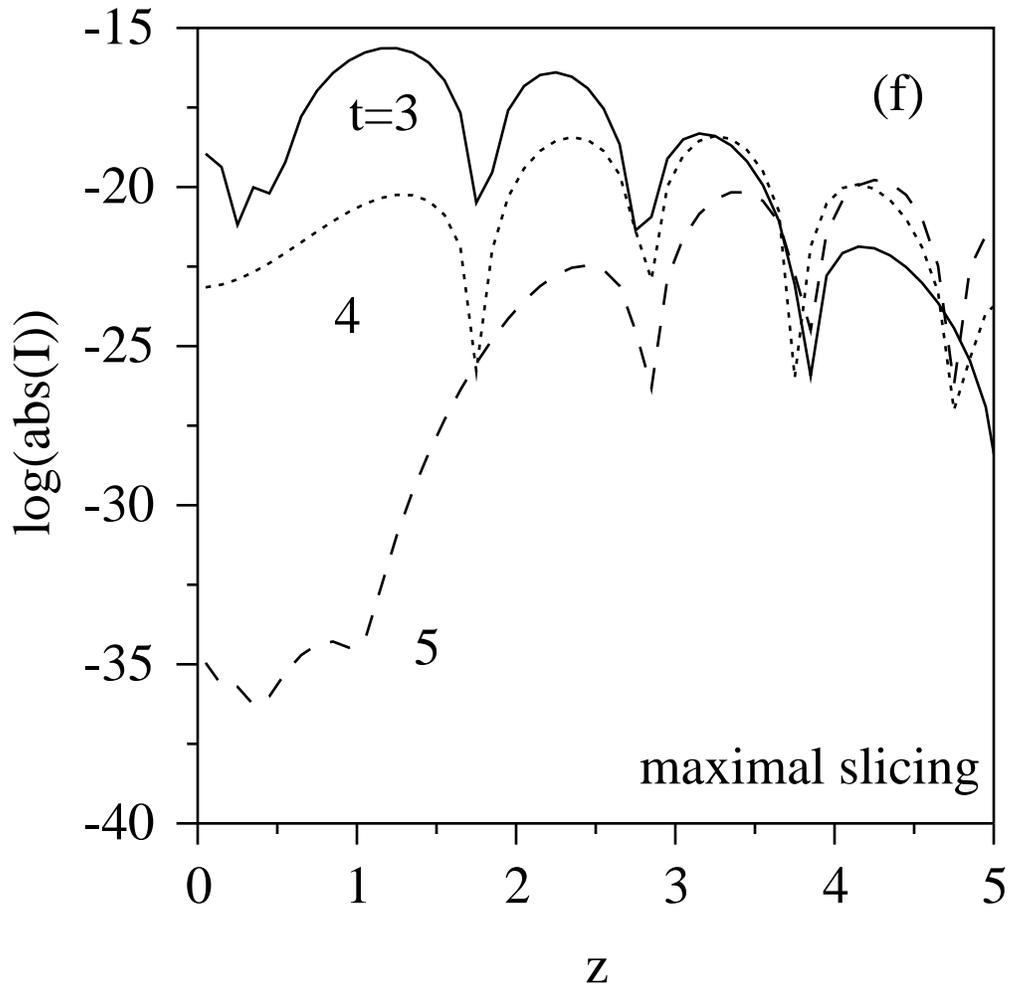
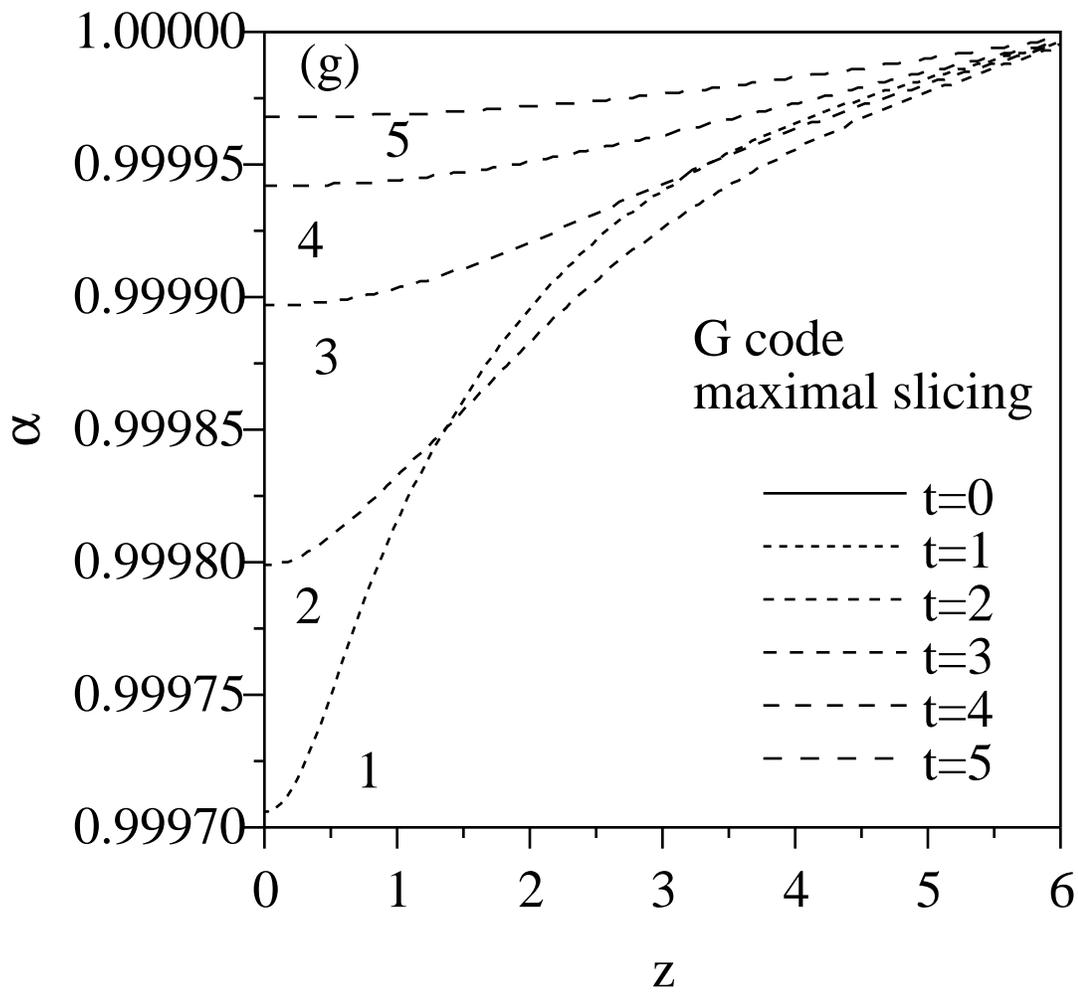
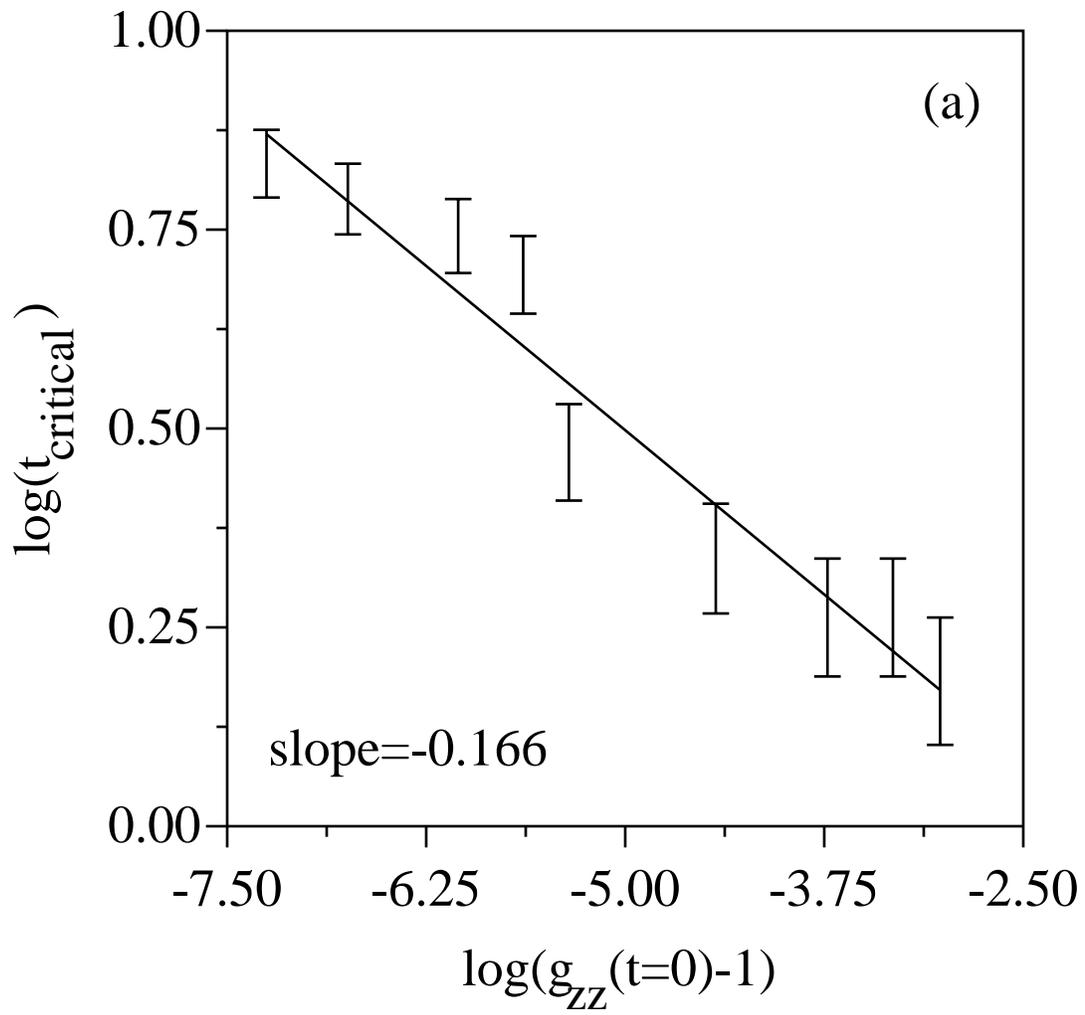
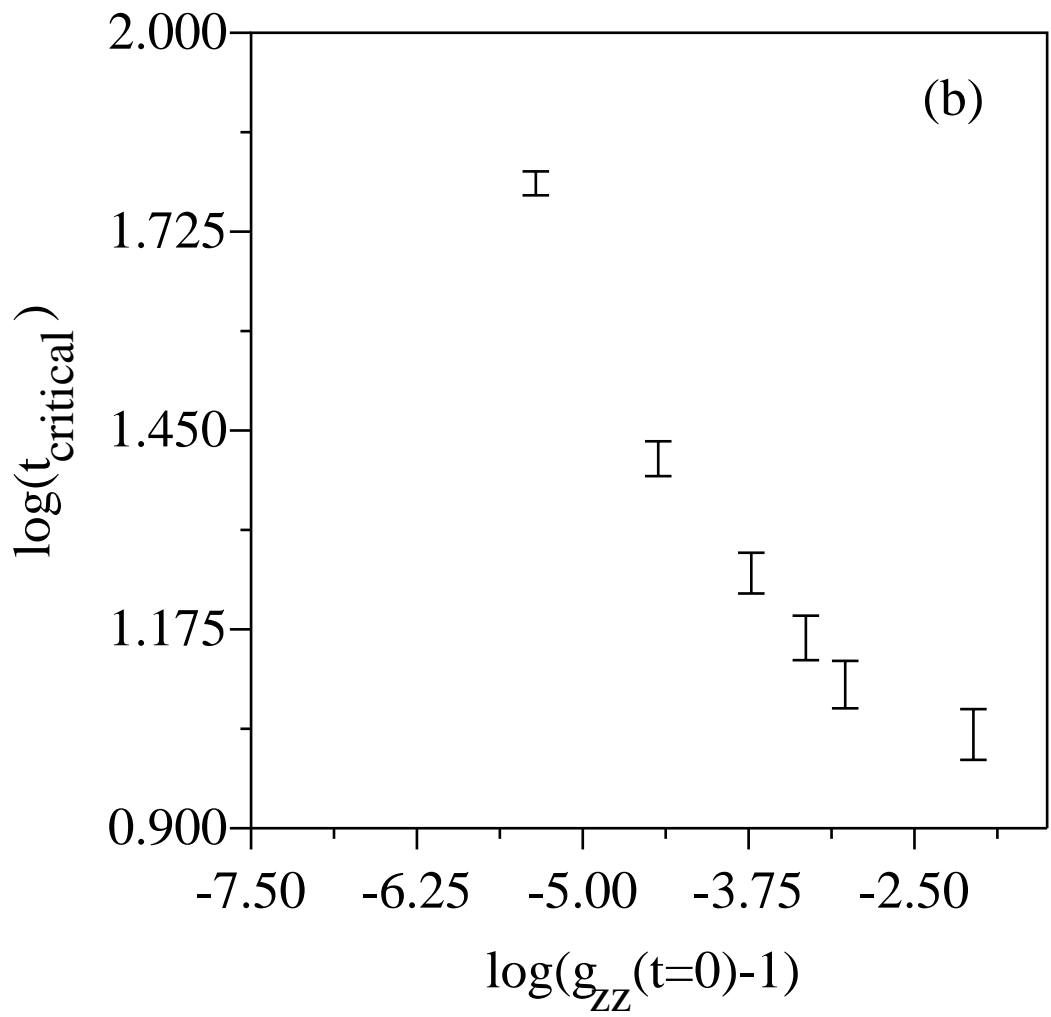
\end{document}